\documentclass[]{aa}       
\usepackage{graphicx}
\usepackage{amssymb}
\usepackage{amsmath}
\usepackage{longtable}
\usepackage{supertabular}
\usepackage{natbib} 

\begin{document}

\title{3D simulations of microquasar jets in clumpy stellar winds}

\author{M. Perucho\inst{1} \and
        V. Bosch-Ramon \inst{2}
    }

\authorrunning{Perucho \& Bosch-Ramon}

\titlerunning{3D simulations of jets in clumpy winds}

\institute{Dept. d'Astronomia i Astrof\'{\i}sica, Universitat de Val\`encia, C/ Dr. Moliner 50, 46100, Burjassot (Val\`encia), Spain; 
Manel.Perucho@uv.es 
\and
Dublin Institute for Advanced Studies, 31 Fitzwilliam Place, Dublin 2, Ireland; valenti@cp.dias.ie
}

\offprints{M. Perucho, \email{manel.perucho@uv.es}}

\date{Received <date> / Accepted <date>}

\abstract
{High-mass microquasars consist of a massive star and a compact object, the latter producing jets that will interact 
with the stellar wind. The evolution of the jets, and ultimately their radiative outcome,
could depend strongly on the inhomogeneity of the wind, which calls for a detailed study.}  
{The hydrodynamics of the interaction between a jet and a clumpy wind is studied, focusing on the global wind- 
and single clump-jet interplay.} 
{We have performed, using the code \textit{Ratpenat}, three-dimensional numerical simulations of a clumpy wind 
interacting with a mildly relativistic jet, 
and of individual clumps penetrating into a jet.}
{For typical wind and jet velocities, filling factors of about $\ga 0.1$ are already enough for the wind to be 
considered as clumpy. An inhomogeneous 
wind makes the jet more unstable when crossing the system. Kinetic luminosities $\sim 10^{37}$~erg/s 
allow the jet to reach the borders of a compact binary with an O star, 
as in the smooth wind case, although with a substantially 
higher degree of disruption. When able to enter into the jet, clumps are compressed 
and heated during a time of about their size divided by the sound speed in the shocked clump. Then, clumps 
quickly disrupt, mass-loading and slowing down the jet.}
{We conclude that moderate wind clumpiness makes already a strong difference with the homogeneous wind case, 
enhancing jet disruption, mass-loading, bending, and likely energy dissipation in the form of emission. All 
this can have observational consequences at high-energies and also in the large scale radio jets.} 

\maketitle

\section{Introduction} \label{intro}

Microquasars are binary systems hosting a star and an accreting black hole or neutron star. Matter from the star  is transferred to the compact 
object, part of it being launched through magnetocentrifugal forces \citep[e.g.,][]{bla77,bla82,bar11}.  This triggers the formation of bi-polar 
jets, which generate non-thermal radio emission \citep[e.g.,][]{mir99,rib05}, and are thought to be the location from where the gamma rays observed 
in some  sources are emitted \citep[e.g.,][]{alb07,tav09,abd09,sab10}. Jets could be magnetically dominated at their base, but 
magnetohydrodynamical processes occurring at higher jet height would accelerate the flow,  efficiently converting magnetic energy into kinetic one 
\citep[e.g.][]{kom07}.  At the scales of the binary system ($\sim 10^6\,R_{\rm Sch}$, where $R_{\rm Sch}$ is the Schwarzschild radius), the jet is 
likely to be already a hydrodynamical (HD) flow. We focus here on the persistent jets thought to be present during the low-hard state of 
microquasars, although some considerations for transient ejecta, associated to low-hard to high-soft state transitions, are done below 
\citep[see, e.g.,][for reviews on microquasar states]{fen04,fen09}.

Part of the energy carried by the jet can be dissipated in the form of magnetic reconnection, recollimation and 
internal shocks, shear layers in the jet walls, and turbulence. Part of the dissipated energy can go to 
non-thermal particles, generating low- and high-energy emission via different mechanisms, 
synchrotron from radio to X-rays, and inverse Compton (IC) and
hadron-related processes up to gamma rays \citep[see, e.g.,][and references therein]{bos09}. 

As shown by \cite{per08} (PB08 hereafter) and \cite{per10} (PBK10 hereafter), in microquasars hosting an OB star 
(high-mass microquasars; hereafter HMMQ) the jet may be strongly influenced by the stellar wind.
The one-side impact of the wind on the (presumably) already HD jet leads to strong and asymmetric recollimation 
shocks, bending and different types of instabilities: the recollimation
shocks seem suitable candidates for particle acceleration and non-thermal emission; bending may be noticeable in 
radio at milliarcsecond scales; instabilities may destroy the jet flow even
within the binary system. For typical wind and jet velocities, say $v_{\rm w}\sim 2\times 10^8$ and $v_{\rm j}\sim 10^{10}$~cm/s, respectively, 
studies show that for compact binaries
and jet-to-wind momentum flux ratios $\la 0.1$ the jet can be already disrupted (PB08, PBK10). This number is 
rather constraining, since only a few HMMQ 
might be above this threshold. This could be the reason for the low number of HMMQ detected, as 
suggested in PBK10. In any case, even if not destroyed within the binary system, jets can suffer strong 
perturbations with both dynamical and radiative consequences. 

Previous work in HMMQ wind-jet interactions was done under the assumption that the wind is homogeneous, but in 
fact stellar winds are thought to be clumpy \citep[e.g.,][]{owo06,mof08}. For
this reason, it has been proposed that wind clumpiness should be taken into account when studying HMMQ 
(\citealt{owo09,ara09} -ABR09 hereafter-; \citealt{rom10,ara11}).  An important parameter that determines the
inhomogeneity of the wind is the wind filling factor $f$, which determines the wind volume fraction with higher 
density. For a significant departure from homogeneity, the intraclump medium
mass and momentum fluxes will be negligible and only clumps will have a dynamical impact on the jet. Since the 
interaction between a HD jet and a clumpy wind has not been studied in detail,
we have carried out 3-dimensional (3D) simulations of this scenario. Simulations have been done for two different 
jet powers and jet-to-wind momentum ratio, $L_{\rm j}=3\times
10^{36}-10^{37}$~erg/s and $\approx 0.03-0.08$, respectively, to explore what could be the transition 
between jet destruction and long-term collimation. Another simulation has
focused on the evolution of individual clumps injected in the jet at different heights. Unlike in PBK10, in which 
the simulation started with the jet being injected at its base, here the
jet is conical and crossing the whole grid, and the clumpy wind is injected from one of the jet sides.

The first goal of this work is the study of the hydrodynamical evolution of a jet when the wind interacting 
with it is clumpy. The second goal is to quantify for which values of the clump and jet
parameters, clumpiness becomes a relevant factor. It is also interesting to study the evolution of a clump under the impact
of a microquasar jet.
The results can also be used to refine radiation models or 
interpret radio observations, although this
will be treated qualitatively. The paper is organized as follows: in Section~\ref{phys}, the scenario studied 
here is briefly introduced; in
Sect.~\ref{sim}, the simulations are described (Sect.~\ref{sim1}); results are shown in 
Sect.~\ref{sim2}; finally, in Sect.~\ref{disc}, the results are discussed in the context of
jet propagation in HMMQ (Sect.~\ref{disc1}), individual clump-jet interactions 
(Sect.~\ref{disc2}), and their implications for the non-thermal emission
(Sect.~\ref{disc3}). Throughout the paper, we will use cgs units. 

\section{Physical scenario}\label{phys}

\begin{figure}[!h]
   \centering
\includegraphics[clip,angle=0,width=\columnwidth]{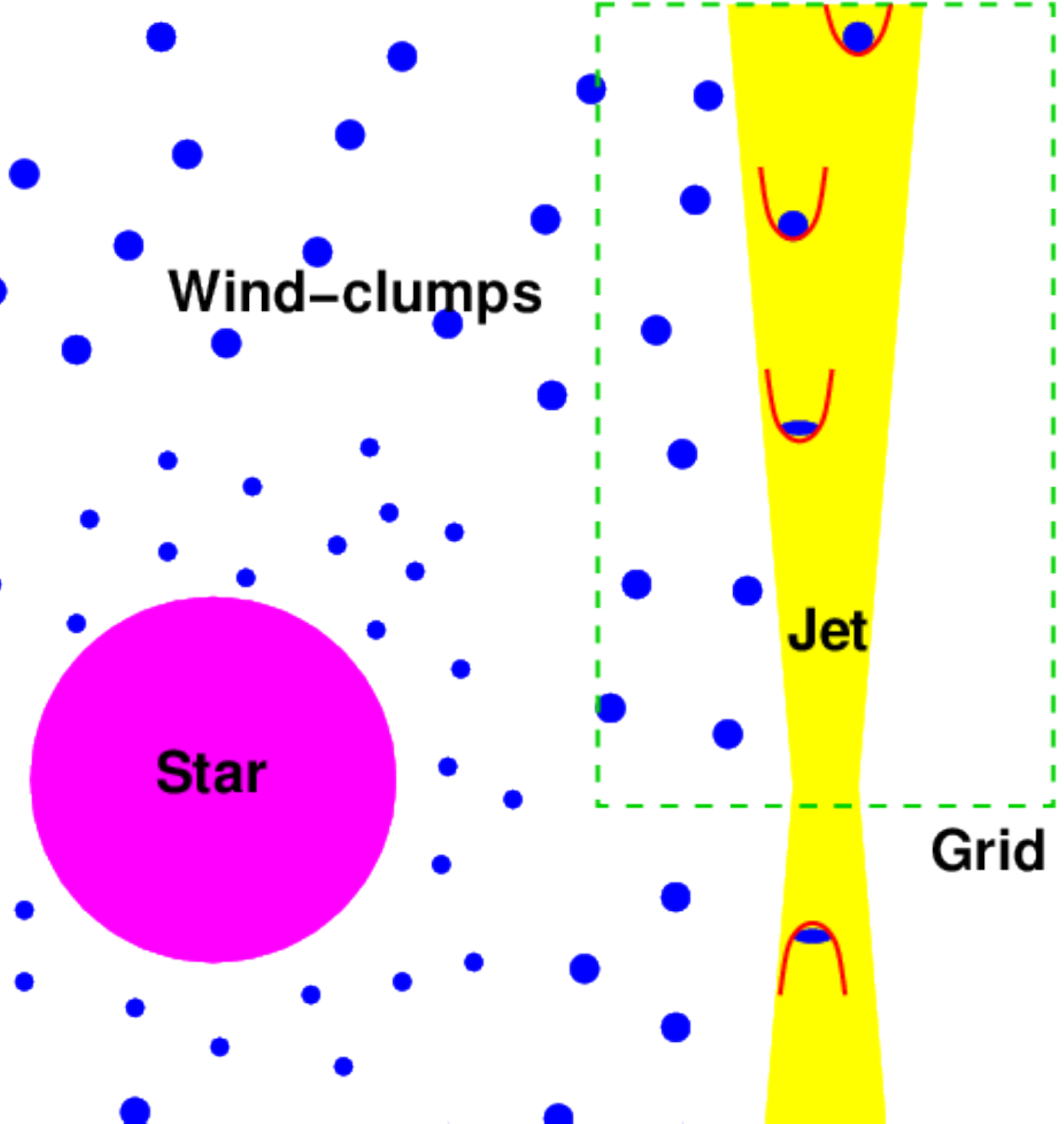}
\caption{Sketch of the scenario simulated in this work. The green dashed line
confines the simulated region.}
\label{fig1}
\end{figure}

The scenario studied here consists of a jet crossing the binary system in a HMMQ. The jet starts close to 
the compact object, which is located at a distance of $d=2\times 10^{12}$~cm
(following PBK10) from the massive star, and is perpendicular to the orbital plane. The jet is initially conical, 
with a radius to height ratio of $\eta=R_{\rm j}/z=0.1$. The scenario is
similar to that studied in PBK10. However, unlike in that work, the stellar wind is assumed here to be 
inhomogeneous, with a filling factor $f\sim 0.1$. A sketch of the considered scenario is
presented in Fig.~\ref{fig1}.

 The inhomogeneities or clumps are modelled as gaussians with $\sigma=R_{\rm c}=3\times 10^{10}$~cm \footnote{The effective size of such 
clumps, if they were spherical and homogeneous, would be $\sim 4\times 10^{10}$~cm.}. In reality, the size, mass, and velocity of clumps in stellar 
winds may follow complex distributions \citep[e.g.,][]{mof08}. However, we have adopted average and representative values for these quantities for 
simplicity. Bigger and thus less numerous clumps than assumed here would probably have a stronger impact on the jet dynamics,
unless they were so few that interactions were rare. The latter may be the case, provided that wind mass seems to concentrate in the small clumps. 
These small clumps may also be denser than big ones \citep{mof08}. On the other hand, if $R_{\rm c}\ll 3\times 10^{10}\,{\rm cm}$, then the wind 
could be effectively considered as homogeneous (as in PBK10). The reason is that clumps
cover inside the jet a fraction of its radius:  $\chi=R_{\rm c}/R_{\rm j}\sim v_{\rm w}\,R_{\rm c}\,(\pi\,v_{\rm j}\rho_{\rm c}/2L_{\rm j})^{1/2}$ 
(for a Newtonian jet), where $\rho_{\rm c}=\dot{M}/4\pi f d^2  v_{\rm w}$ is the clump density. This estimate is based on the clump disruption time, 
expected to be slightly longer than the clump shock crossing time: $t_{\rm d}\sim
R_{\rm c}/c_{\rm c}$, where  $c_{\rm c}\sim(2\,L_{\rm j}/\pi v_{\rm j}\rho_{\rm c}R_{\rm j}^2)^{1/2}$ is the shocked clump sound speed. 

Another relevant timescale is the clump acceleration time along the jet, i.e., the time needed to accelerate the clump material up to a speed 
$\sim c_{\rm c}$, which is $\sim t_{\rm d}$, and because of quick expansion, it takes only several times this value for the clump material to 
reach $\sim v_{\rm j}$ (see, e.g., \citealt{bla79}; \citealt{kle94}; ABR09; \citealt{pit10}; \citealt{bar10}; for shocked clump evolution in different 
contexts). For stellar mass-loss rates $\dot{M}\sim 10^{-6}\,M_\odot$/yr, $L_{\rm j}\sim 10^{37}$~erg/s, $d\sim 2\times 10^{12}$~cm, and 
$v_{\rm j}\sim 10^{10}$~cm/s, $\chi$ is 
$\sim 0.1\,(R_{\rm c}/3\times 10^{10}\,{\rm cm})\,(f/0.1)^{-1/2}$, i.e., clumps  with $R_{\rm c}\ll 3\times 10^{10}\,(f/0.1)\,{\rm cm}$ are destroyed 
in the external jet layers. In this small clump case the wind-jet contact 
discontinuity should probably develop a turbulent shear-layer faster than in the homogeneous case. Due to destruction, therefore, clumps will not be 
able to enter the jet if $c_{\rm c}<v_{\rm w}$ within the binary, which occurs at $f>0.1$ for the parameters given above. For relatively weak jets, 
say $L_{\rm j}\lesssim 10^{36}$~erg/s, these clumps can otherwise cross the whole jet. For $L_{\rm
j}\sim 10^{37}$~erg/s, the clumps considered may enter the jet, but it would be difficult for them to escape (see Sect.~\ref{disc3}).

From the above discussion, we conclude that, under the jet-to-wind momentum flux ratios considered, if the wind is moderately inhomogeneous, say 
$f\lesssim 0.1$, clumps can effectively penetrate and mass-load
the jet, with a penetration effectiveness depending on $R_{\rm c}$. As we show in what follows, clump penetration will have serious consequences on 
the jet stability and long-term collimation. We note that we have assumed that the properties of the clumps, and the  mean separation between them, 
are constant all over the grid region in which they are located for simplicity. This approximation starts to fail for $z\gtrsim d$, although most of 
the dynamical impact occurs below or around this distance. It is worthy to mention that for the purpose of
this work, the assumed hydrodynamical nature of the jet could be loose to some extent, as long as the magnetic to kinetic pressure ratio in the jet 
is smaller, say, than $\sim 0.1$.
Otherwise, the lateral magnetic pressure can effectively prevent clump penetration into the jet. Moreover, assuming that clumps managed to enter the 
jet, magnetic pressure could suppress the development of shocks, induce magnetic dissipation and non-thermal activity, and change strongly the clump 
disruption process, softening or potentiating it
\citep[e.g.][]{jon96,shi08}. In any case, the flow evolution under a dynamically dominant magnetic field is out of the scope of this work.

\section{Simulations}\label{sim}
We have performed three numerical simulations using a finite-difference code named \textit{Ratpenat}, which 
solves  the equations of relativistic hydrodynamics in three dimensions,
written in conservation form, using  high-resolution-shock-capturing methods. \textit{Ratpenat} was parallelized 
with a hybrid scheme with both  parallel processes (MPI) and parallel
threads (OpenMP) inside each process \citep[see][]{pe10}. The  simulations were performed in Mare Nostrum, 
at the Barcelona Supercomputing Centre (BSC) with 200 processors, each of
them with a duration of 1080~hours, amounting a total of $6.48\times10^5$ computational hours.

\subsection{Simulation set-up}\label{sim1}

The simulations are set-up with an overpressured jet surrounded by the stellar wind of the massive companion. We 
are thus implicitly assuming that the bow shock generated by the jet when
crossing this medium (PBK10) is far enough and the cocoon dilute enough that the wind has occupied the space 
surrounding the jet itself \footnote{The system we simulate is dynamic so the
initial conditions are artificial. However, the long-term qualitative result is not affected by this, as the 
{\it steady} jet is not in equilibrium but evolves governed by the interaction
with the wind.}. The physical size of the grid is, in (base) jet radius units, 
$160\, R_{\rm j} \, \times \, 160 \, R_{\rm j} \, \times \,200\,R_{\rm j}$, the last $200\,R_{\rm j}$ in the direction of propagation of 
the
jet ($z$ coordinate). The grid size in the $x$ and $y$ coordinates is divided in two regions, the inner $80\,R_{\rm j}$ 
around the jet axis having homogeneous resolution, and the outer $40\,R_{\rm j}$
in each direction being formed by cells with increasing size. The resolution in the homogeneous grid is 
4~cells/$R_{\rm j}$ at injection\footnote{Note that the jet radius increases with
$z$, and so the effective resolution across the jet does.}, with a total of $320 \times 320 \times 800$ cells. 
The  extended grid is composed by 80~cells on each side, resulting in a box
with $480\times480\times800$ cells. In the simulations, $R_{\rm j}=2\times10^{10}\,\mathrm{cm}$, so the physical size 
of the grid is $(3.2\times3.2\times4)\,\times \,10^{12}$~cm. 

\begin{figure*}[!t]
    \centering
  \includegraphics[clip,angle=0,width=0.9\textwidth]{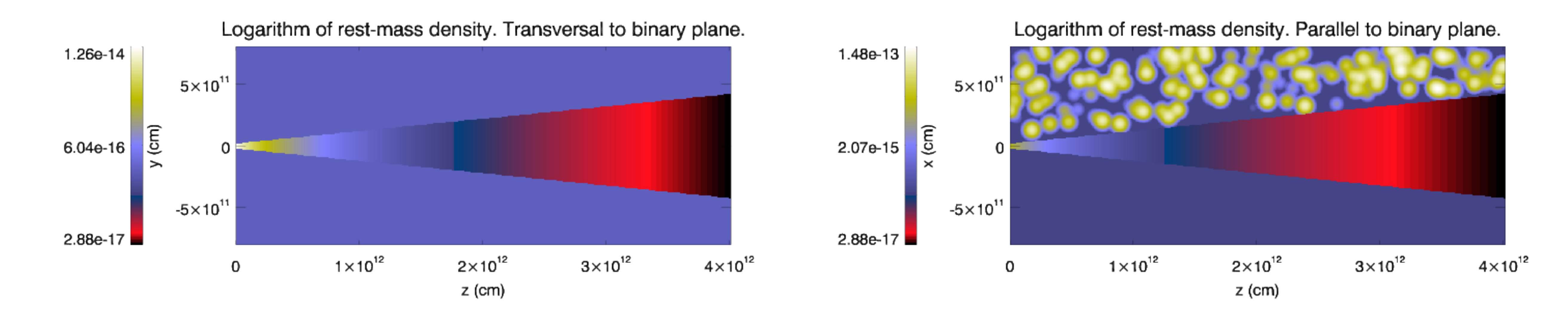}
  \includegraphics[clip,angle=0,width=0.9\textwidth]{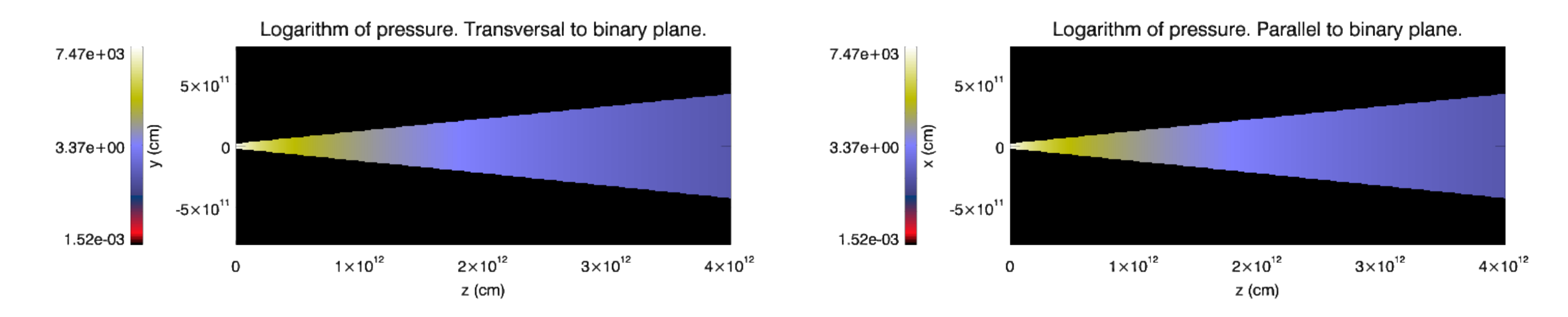}
  \includegraphics[clip,angle=0,width=0.9\textwidth]{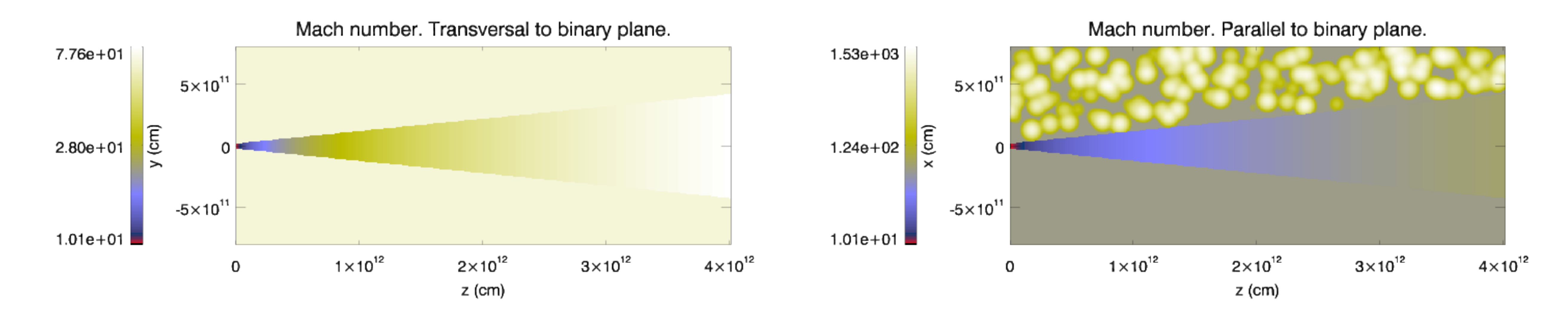}
  \caption{Axial cuts of the jet in the planes perpendicular (left) and parallel (right) to the jet-star plane of rest-mass 
density (top, $\mathrm{g/cm^{3}}$), pressure (mid, $\mathrm{erg/cm^{3}}$) and Mach number (bottom) for one of the initial 
models.}
  \label{fig:maps1}
  \end{figure*}

\begin{figure}[!h]
    \centering
  \includegraphics[clip,angle=0,width=\columnwidth]{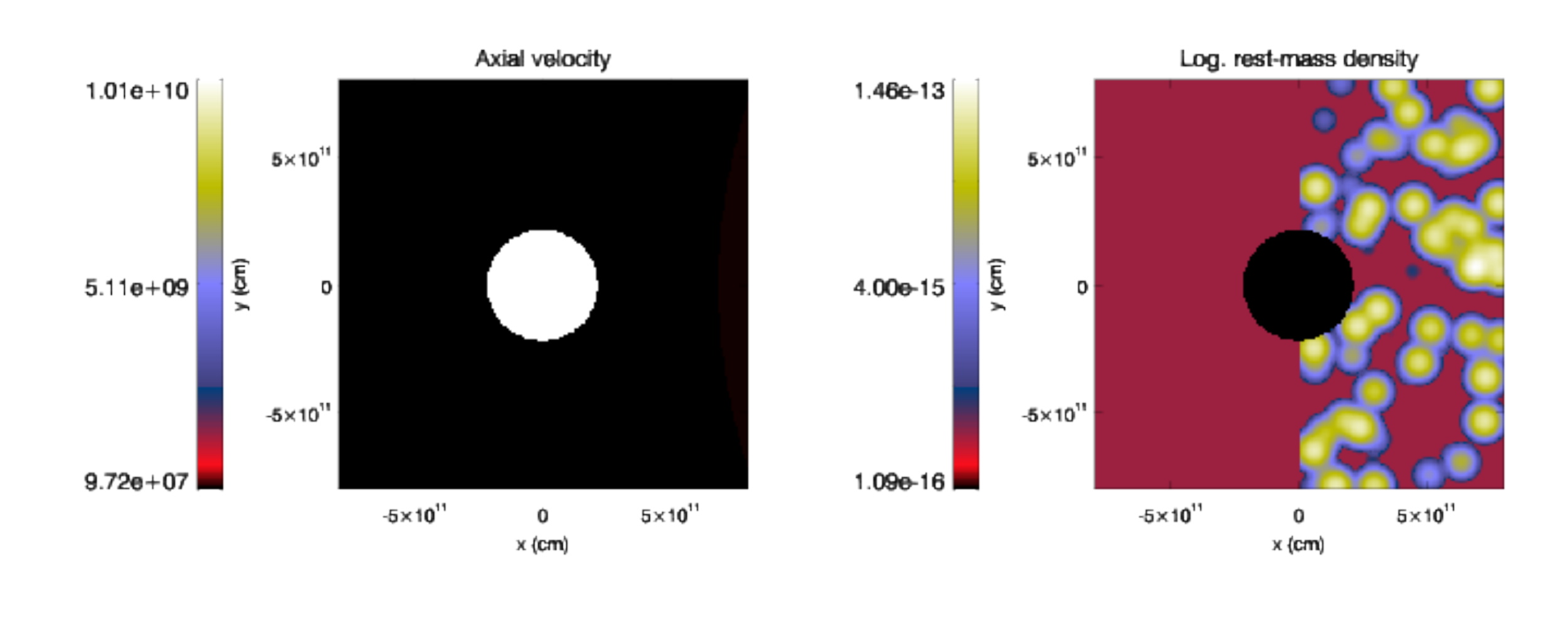}
  \caption{Transversal cuts of the jet at half grid of axial velocity ($\mathrm{cm/s}$) and rest-mass density 
($\mathrm{g/cm^{3}}$) for one of the initial models.}
  \label{fig:maps2}
  \end{figure}

In previous works (PB08, PBK10), the wind was simulated as homogeneous, but it was already suggested that a 
necessary improvement to those simulations should be the inclusion of
inhomogeneities. We have done this by randomly adding Gaussian-shaped clumps in the side of the grid from where 
the stellar wind is injected. The initial number of clumps in the half of the
homogeneous grid facing the star, 1500, has been calculated to ensure that the mean wind mass-loss rate is 
$\approx 10^{-6}\,M_\odot$/yr. The clump peak density is 10 times larger than the mean
density ($9\times10^{-15}\mathrm{g/cm^3}$), i.e., $9\times10^{-14}\mathrm{g/cm^3}$, and the minimum density 
between clumps has been fixed to $3\times10^{-16}\mathrm{g/cm^3}$. The whole wind region is
set in pressure equilibrium  ($P_{\rm w}=1.5\times10^{-3}~\mathrm{erg/cm^3}$) with the higher density regions, 
and a velocity $v_{\rm w}=2\times10^8~\mathrm{cm/s}$ radial from the star (PB08, PBK10).
Figure~\ref{fig:maps1} shows cuts of the initial conditions in pressure and density along the axis of one of the 
simulated jets, and Fig.~\ref{fig:maps2} shows transversal cuts at $z\simeq2.2\times10^{12}\,\mathrm{cm}$.

We have simulated two jets with different powers, jet A with $L_{\rm j}=3\times10^{36}\,\mathrm{erg/s}$ and jet B 
with $L_{\rm j}=10^{37}\,\mathrm{erg/s}$ (to be compared with jet 2 in PBK10).
The jets are set-up with a given opening angle but not in pressure equilibrium with the ambient, as this has 
homogeneous pressure whereas the jet pressure decreases with $z$. The injection
point in the grid would be located at $2\times10^{11}~\mathrm{cm}$ from the compact object, this is, well within 
the binary system. The jets have both injection densities of
$4.2\times10^{-15}\mathrm{g/cm^3}$ and $1.4\times10^{-14}\mathrm{g/cm^3}$, and a velocity 
$v_{\rm j}=10^{10}\,\mathrm{cm/s}$. Thermal cooling terms, following the approximation used in
\cite{mya98}, have been added to the code to account for the cooling in the clumps. Table~\ref{tab} summarizes 
the jet and wind parameters used in the simulations.   

We note that to simulate continuous wind injection, the up-wind section of the ambient medium is replenished with 
clumps when a portion of its volume has been emptied of the original ones.
This is done without any effect on the dynamics of the system, as the clumps are added in a region far from the 
interaction between the jet and the first clumps.

\begin{table*}{
\label{tab}
\begin{center} 
  \begin{tabular}{lccccc}
    \hline
     &  Density (g/cm$^3$)  & Velocity (cm/s) & Mach number & Pressure (erg/cm$^3$) & Power (erg/s) \\
     \hline
  Jet A (A$^\prime$)  & $4.2\times10^{-15}$ & $10^{10}$ & $10$  &  $2274$   & $3\times10^{36}$ \\
  Jet B  & $1.4\times10^{-14}$ & $10^{10}$ & $10$ &  $7470$   & $10^{37}$  \\
  \hline
    &  Min. density (g/cm$^3$)  &  Max. density(g/cm$^3$) & Mean density (g/cm$^3$) & Pressure (erg/cm$^3$) & Velocity (cm/s)  \\
 \hline 
  Wind & $3\times10^{-16}$ & $9\times10^{-14}$ &  $9\times10^{-15}$ & $1.5\times10^{-3}$ & $2\times10^8$ \\
  \end{tabular}
\end{center}
\caption{Jet and wind parameters.}
}
\end{table*}

\begin{figure}[!h]
     \centering
  \includegraphics[clip,angle=0,width=\columnwidth]{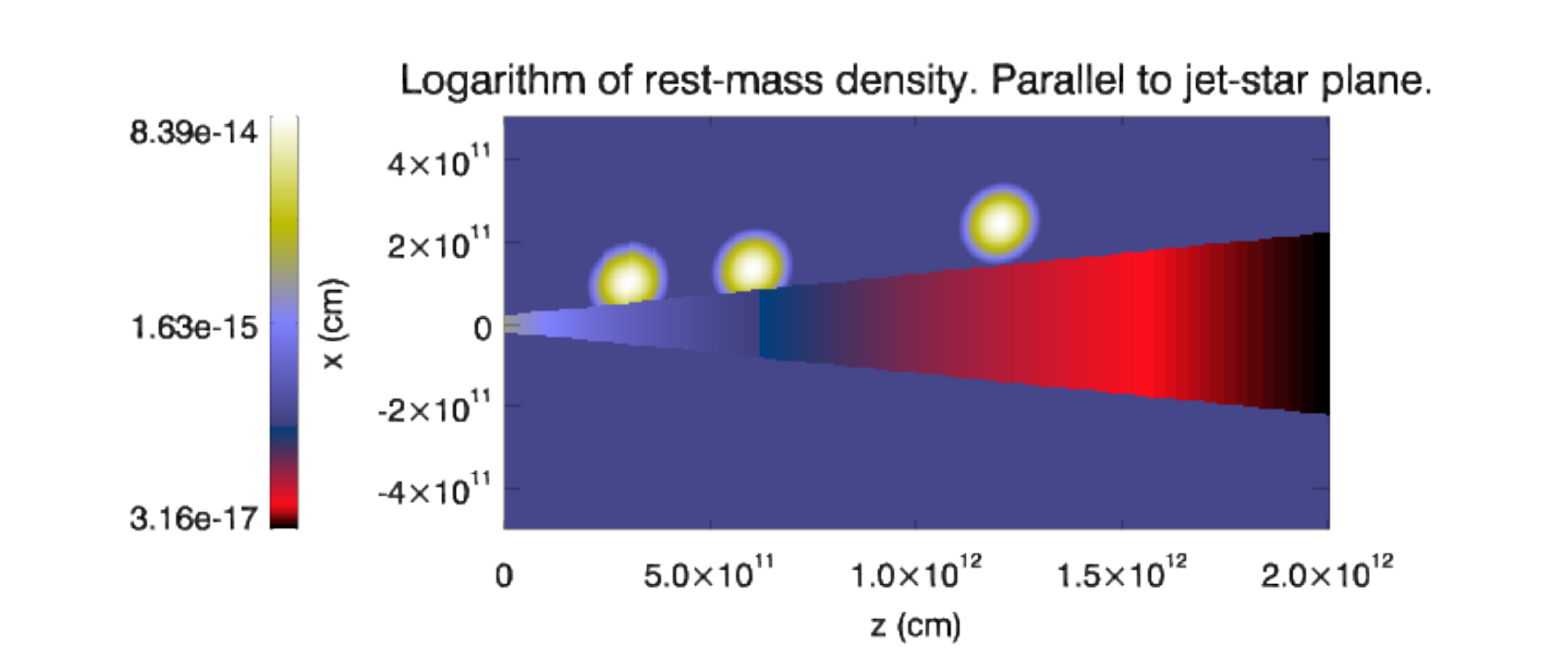}
  \caption{Axial cut parallel to the star-jet plane of rest-mass density ($\mathrm{g/cm^{3}}$) for jet A$^\prime$ showing the 
initial location of the clumps.}
  \label{fig:maps3}
  \end{figure}

A third simulation was performed to illustrate the evolution of 3 clumps interacting with jet~A 
(jet~A$^\prime$). The clumps were located at
$z=3,\,6$~and~$12\times10^{11}\,\mathrm{cm}$, with increasing distance to the jet in the $x$-direction such that 
the different clumps start their interaction with the jet subsequently. In
Figure~\ref{fig:maps3}, the initial distribution of the clumps is shown in the axial density cut of the jet parallel 
to the star-jet plane.

\section{Results}\label{sim2}

Jets initially expand due to their overpressure. The expansion is asymmetric, being faster in the direction of 
propagation of the wind (down-wind). The clumps that enter the jet close to
its base are shocked by the expansion of the overpressured jet and eroded by its transversal velocity. The shock 
generated by this interaction propagates inside the jet, deforming its
surface  and triggering helical displacements of the jet flow, which add to the deviation of the jet in the 
direction of the wind further up. The material ablated from the clumps is
dragged by the jet, circulates around its surface and generates a turbulent thick shear-layer where is 
mixed with material of the jet and accelerated in the direction of the latter. 

In the central region of the jet, a reconfinement shock is triggered by the irregular ram pressure of the clumpy 
wind. At this shock, the jet flow is decelerated, favoring clump penetration and
clump-jet material mixing downstream. At higher values of coordinate $z$, clumps can fully penetrate into 
the jet generating bow shocks that cover large fractions of the jet cross section and further
decelerate the flow. Thus, the nature of the jet is completely changed within the binary system.

\subsection{Jet~A}
 
\begin{figure*}[!t]
    \centering
  \includegraphics[clip,angle=0,width=0.9\textwidth]{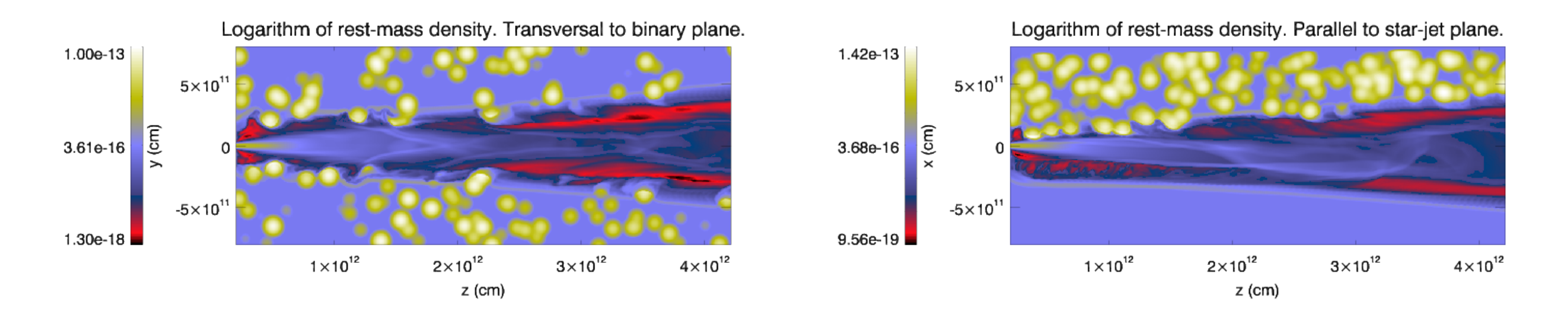}
  \includegraphics[clip,angle=0,width=0.9\textwidth]{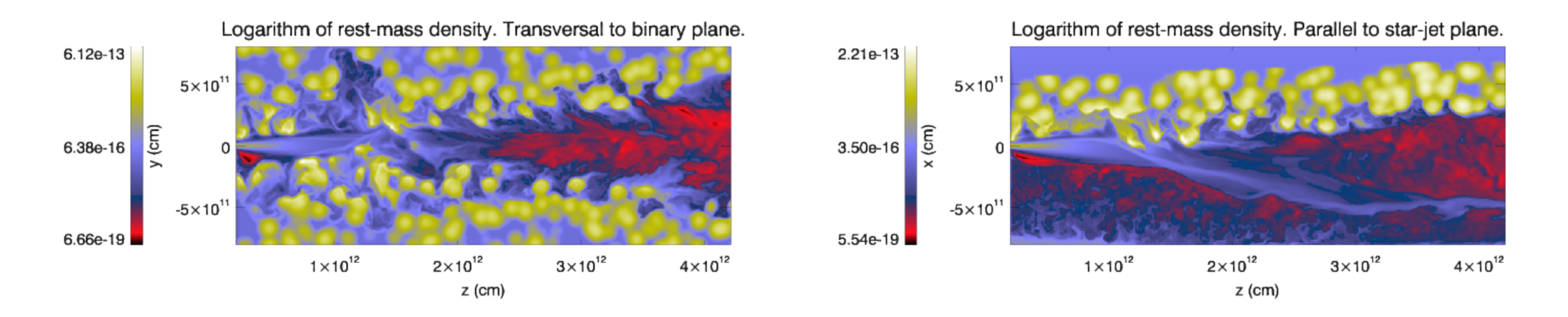}
  \includegraphics[clip,angle=0,width=0.9\textwidth]{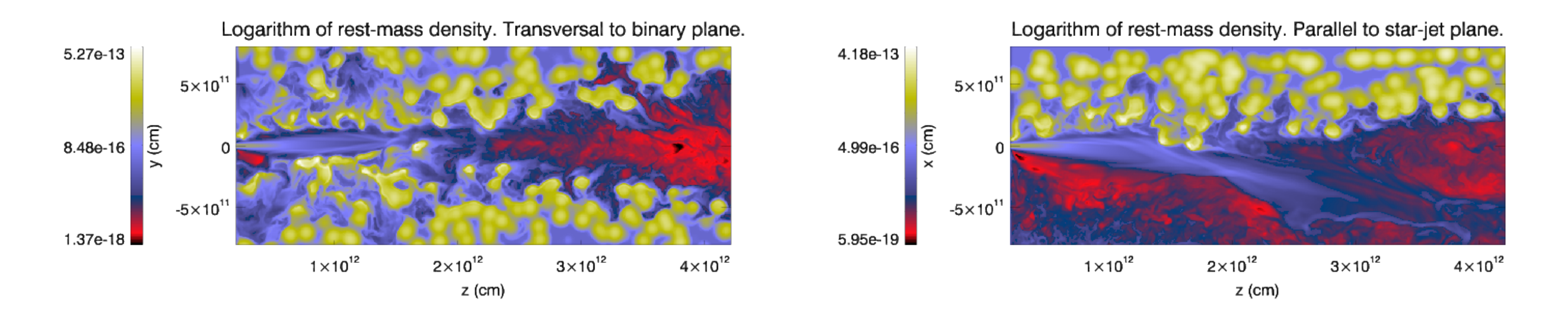}
  \includegraphics[clip,angle=0,width=0.9\textwidth]{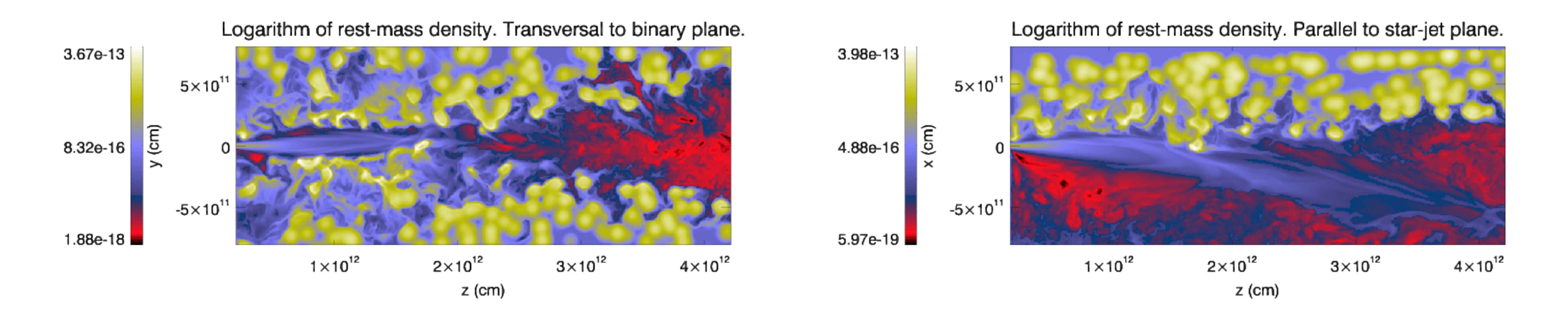}
  \caption{Axial cuts of rest-mass density ($\mathrm{g/cm^{3}}$) at different instants for jet A. Top panels stand for 
$t=300$~s, second row for $t=1220$~s, $t=1570$~s the third one, and bottom one for the last snapshot of the simulation, at 
$t=1700$~s.}
  \label{fig:maps4}
  \end{figure*}

\begin{figure*}[!t]
     \centering
  \includegraphics[clip,angle=0,width=0.9\textwidth]{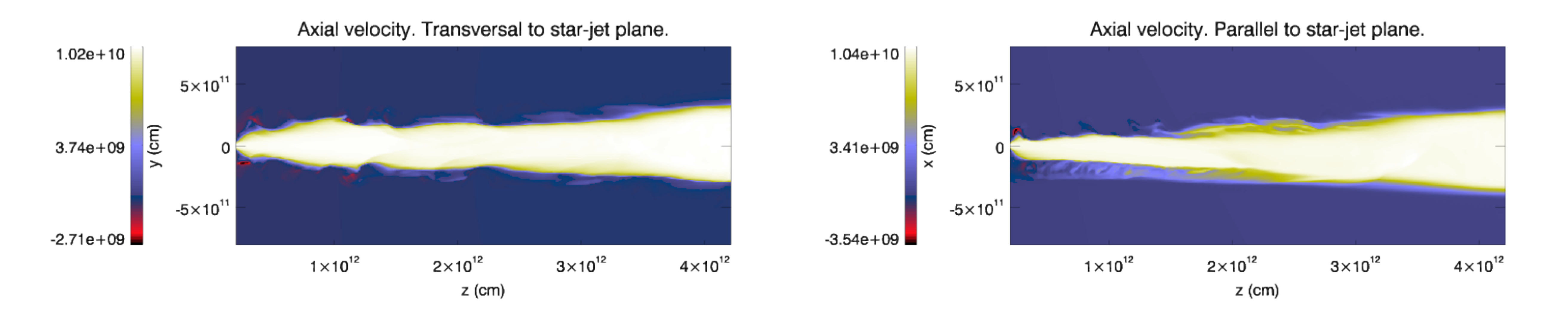}
  \includegraphics[clip,angle=0,width=0.9\textwidth]{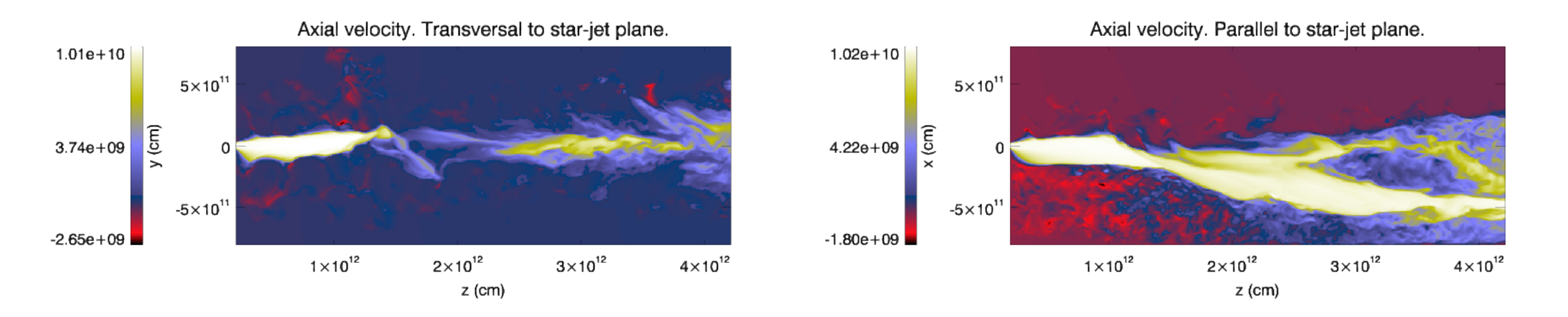}
  \includegraphics[clip,angle=0,width=0.9\textwidth]{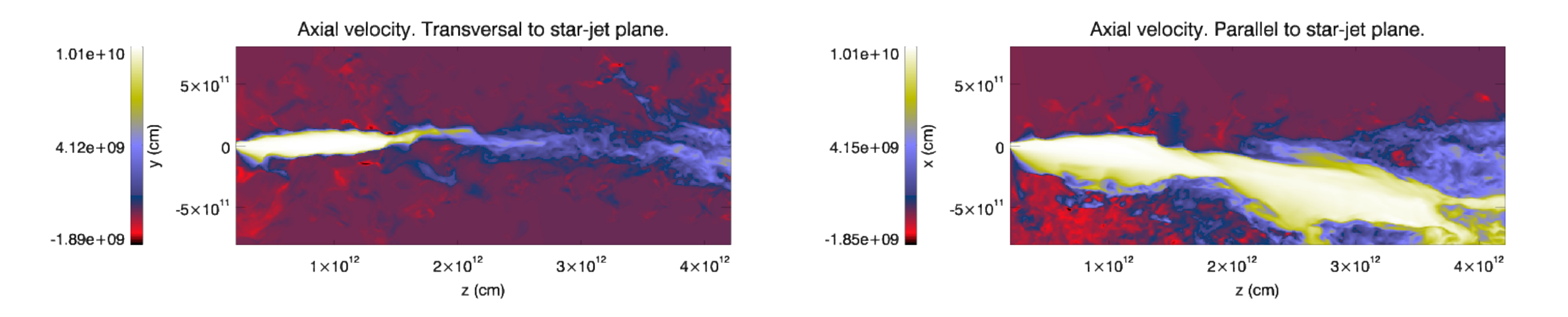}
  \includegraphics[clip,angle=0,width=0.9\textwidth]{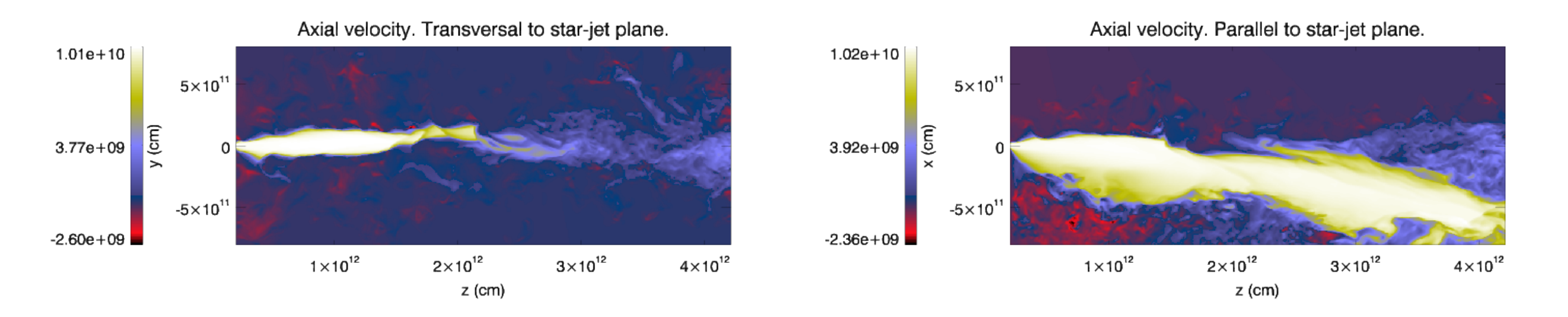} 
  \caption{Axial cuts of axial velocity ($\mathrm{cm/s}$) at different instants for jet A. Top panels stand for $t=300$~s, 
second row for $t=1220$~s, third one for $t=1570$~s, and bottom one for the last snapshot of the simulation, at $t=1700$~s.}
  \label{fig:maps5}
  \end{figure*}

\begin{figure*}[!t]
    \centering
  \includegraphics[clip,angle=0,width=0.9\textwidth]{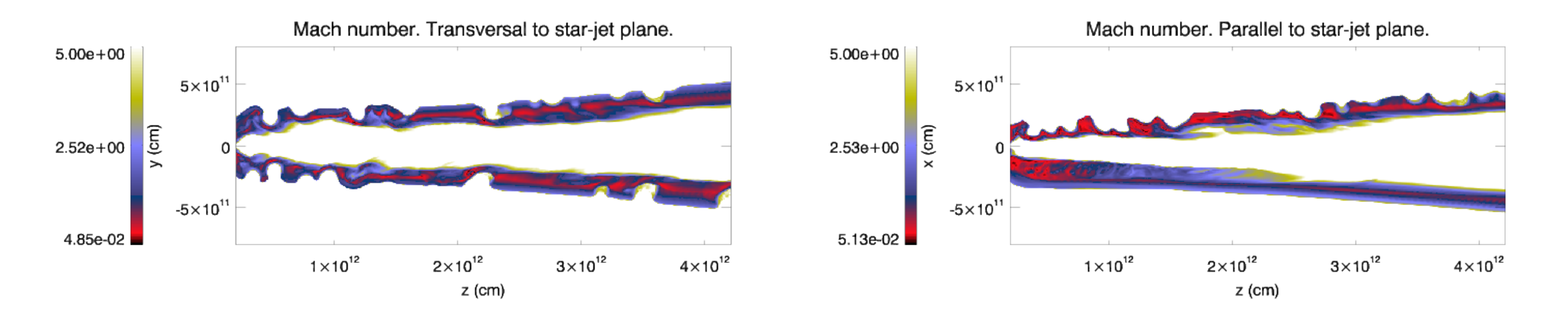}
  \includegraphics[clip,angle=0,width=0.9\textwidth]{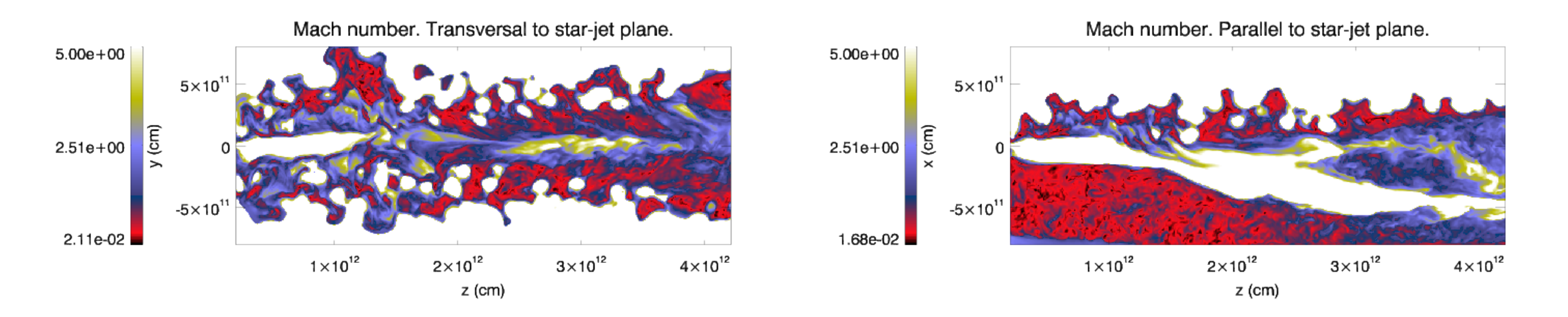}
  \includegraphics[clip,angle=0,width=0.9\textwidth]{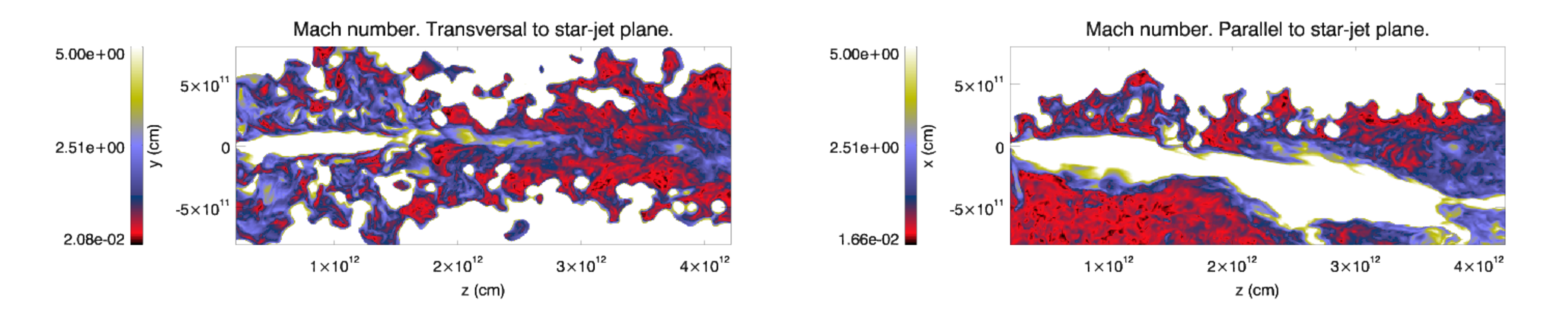}
  \includegraphics[clip,angle=0,width=0.9\textwidth]{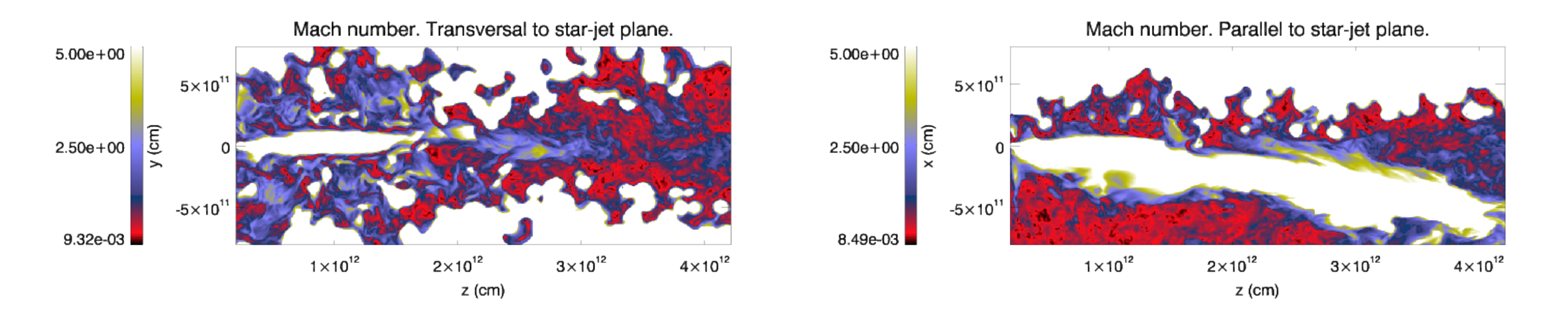}
  \caption{Axial cuts of Mach number at different instants for jet A. Top panels stand for $t=300$~s, second 
row for $t=1220$~s, third one for $t=1570$~s, and bottom one for the last snapshot of the simulation, at $t=1700$~s. The color scale 
for this parameter has been saturated at $M=5$, 
thereby enhancing the contrast inside the jet and thus the location of transonic flows resulting from shocks.}
  \label{fig:maps6}
  \end{figure*}

In the case of jet~A, the simulation reproduces $\simeq 1700$~s of the jet-wind interaction. Due to the high 
overpressure of the jet at the injection point, the clumps are destroyed when they
get close to the flow at the jet base, although much farther up they can penetrate inside the jet. 
Figures~\ref{fig:maps4}, \ref{fig:maps5} and \ref{fig:maps6} show two jet axial cuts,
transversal and parallel to the star-jet plane, of rest-mass density, axial velocity and Mach number, 
respectively. The different images show the temporal  evolution, with time increasing
downwards. The initial phase of the interaction between the clumps and the jet is seen in the upper 
panels of each of the figures. The density maps show the generation of an asymmetric
reconfinement shock (right panel) due to the ram-pressure of the clumpy wind, and the generation of a rarefied 
region in the boundaries of the jet produced by the initial expansion into the
wind diluted regions. In this slower and more dilute rarefaction region, clumps penetrate easily, deforming 
the jet surface (see Fig.~\ref{fig:maps5}), and the up-wind jet expansion
becomes dominated by the interaction with the clumps. 

At the adopted initial conditions, the jet cools down as it expands along the $z$ axis, accelerating also slightly in the 
$z$-direction. This implies an increase of the Mach number from the value at
injection, $M=10$, to $M\simeq 60$. In the Mach number maps, we can see that, as the time goes by, the maximum 
Mach numbers in the jet downstream of the
reconfinement shock get much smaller than the initial values. Deceleration due to mass entrainement, and heating due to shocks, lead 
to the formation of transonic 
and subsonic regions inside the jet flow.

The bottom panels of Figs.~\ref{fig:maps4}, \ref{fig:maps5} and \ref{fig:maps6} show the last snapshot of the 
simulation of jet~A. The cuts transversal to the star-jet plane (left panels) reveal the jet
expansion at small $z$, reconfinement and, once the jet has been deviated, the turbulent region where clump-jet 
mixing and jet deceleration occur. The cuts parallel to the star-jet plane
(right panels) show the effect that a clump entrained just after the reconfinement shock may have, triggering a 
deviation in the jet direction of more than $10^{\circ}$. As noted above, in
the transversal cut to the star-jet plane of rest-mass density (Fig.~\ref{fig:maps4}, bottom left  panel), it can 
be seen that the clumps approaching the jet close to the injection point,
where the overpressure is large, are destroyed before entering it. In this region, the figures also show complex 
structures of destroyed clumps mixed with jet material.

\begin{figure*}[!t]
    \centering
  \includegraphics[clip,angle=0,width=\textwidth]{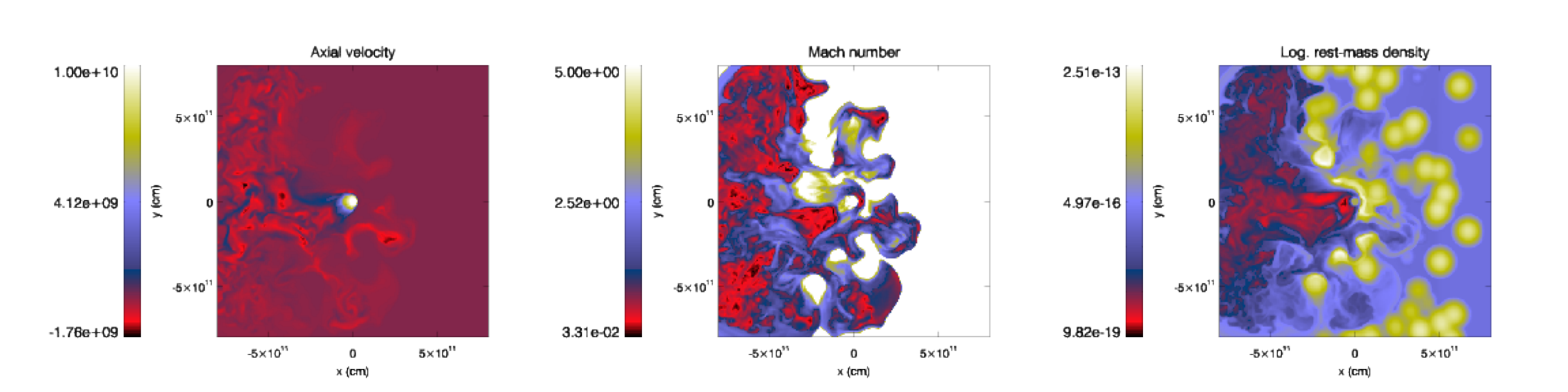}
  \includegraphics[clip,angle=0,width=\textwidth]{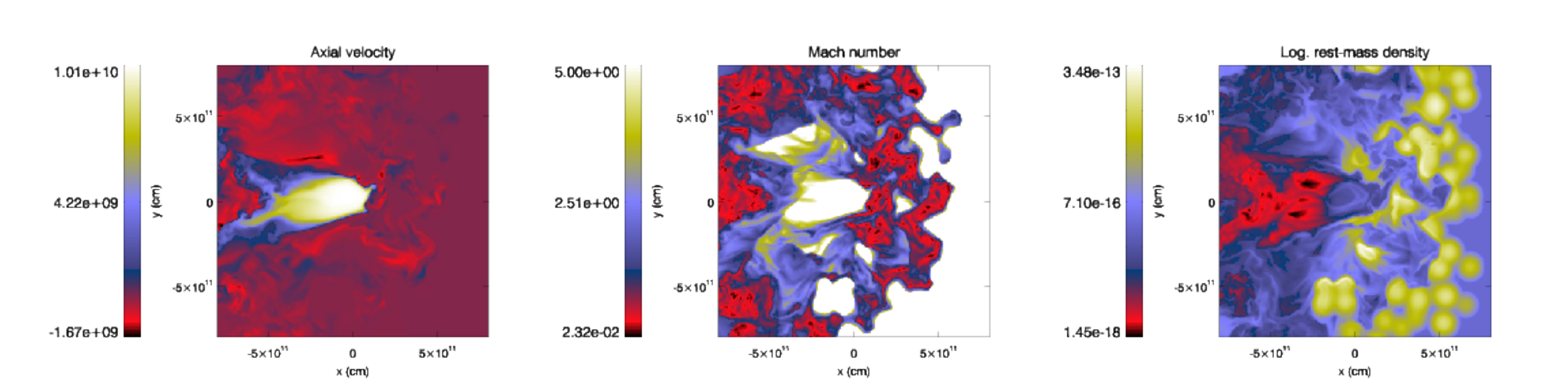}
  \includegraphics[clip,angle=0,width=\textwidth]{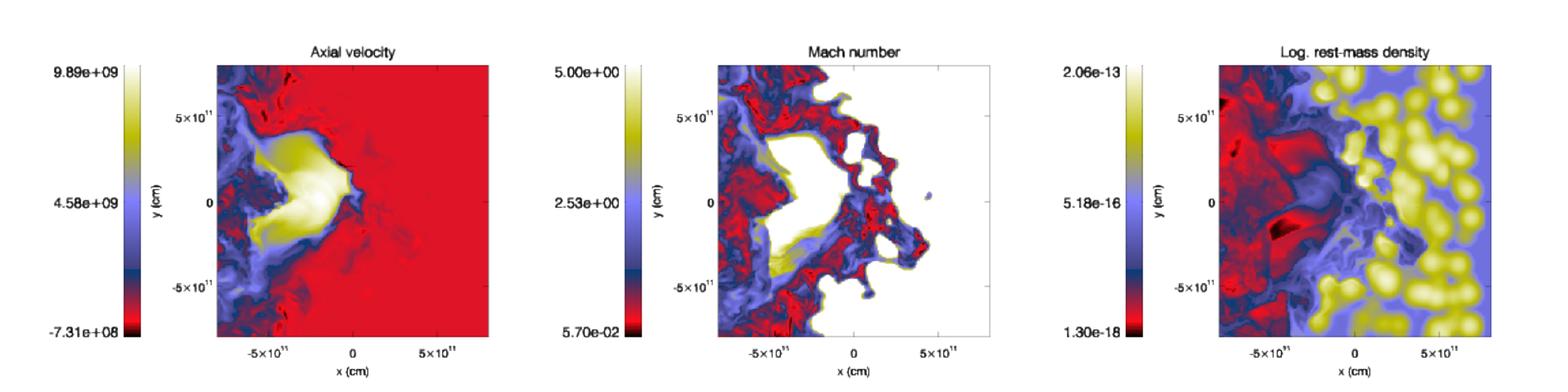}
  \includegraphics[clip,angle=0,width=\textwidth]{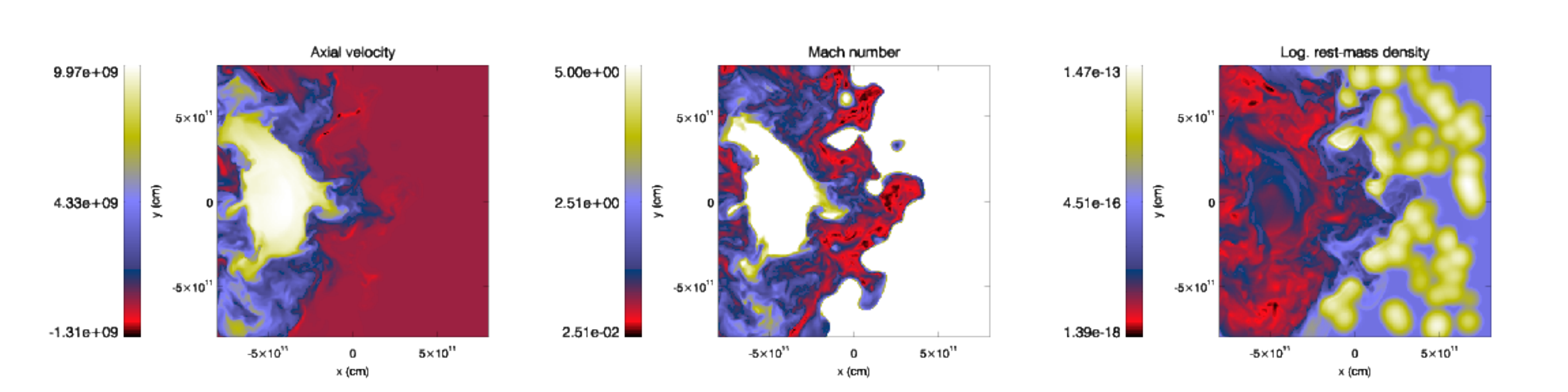}
  \caption{Transversal cuts of rest-mass density ($\mathrm{g/cm^3}$), axial velocity ($\mathrm{cm/s}$) and Mach number for the 
last snapshot ($t=1700$~s) of jet A, at 
$z=2\times10^{11},\,1.2,\,2.2,$~and~$3.2\times 10^{12}\,\mathrm{cm}$. The color scale of the Mach number has 
been saturated at $M=5$.}
  \label{fig:maps7}
  \end{figure*}

Figure~\ref{fig:maps7} shows cuts of rest-mass density, axial velocity and Mach number transversal to the jet axis for jet A at the 
injection point, $\sim 1/4,\, 1/2$, and $3/4$ of the grid length ($z=2\times10^{11},\,1.2,\,2.2,$~and~$3.2\times
10^{12}\,\mathrm{cm}$). The first cut shows the disruption of the clumps as soon as they get close to the jet, 
whereas the jet Mach number and velocity keep their initial values. At one
quarter of the grid ($z\simeq 1.2\times10^{12}\,\mathrm{cm}$),  the jet presents an elongated structure in the 
down-wind direction. In the up-wind direction, rarefied jet material left after the
initial jet expansion forms filaments between the clumps. At this 
point, some regions of the jet section keep a large fraction of the original Mach number and
axial velocity, but some others already show signs of the action of the reconfinement shock. 
At $z\simeq 2.2\times 10^{12}\,\mathrm{cm}$, the jet flow has already gone through this 
shock, and the wind thrust has managed to deviate its original direction. Although the flow has reexpanded 
and partially recovered the strongly supersonic nature, the impact of the reconfinement shock has caused a noticeable decrease in 
velocity from the original values.
At $z\simeq 3.2\times 10^{12}\,\mathrm{cm}$ the jet shows a large radial displacement and
irregular morphology. Overall, the jet section is clearly displaced from the center of the grid. Although the jet 
keeps a high axial velocity at this position, the density cuts show an increase of more than one order of 
magnitude with respect to the original value ($\simeq 3.2\times 10^{-17}\,\mathrm{g/cm^{-3}}$) in some regions of 
the jet at this position.


 \subsection{Jet~B}

\begin{figure*}[!t]
    \centering
  \includegraphics[clip,angle=0,width=0.9\textwidth]{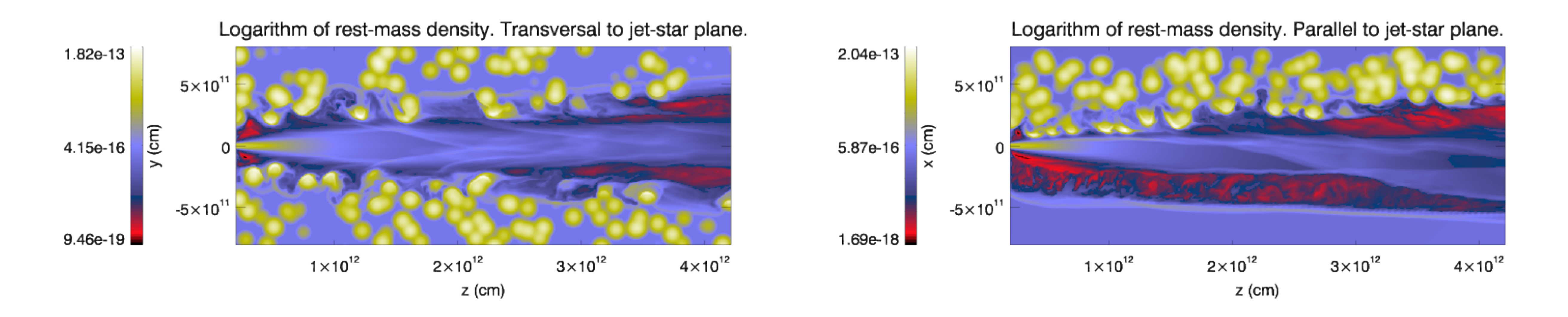}
  \includegraphics[clip,angle=0,width=0.9\textwidth]{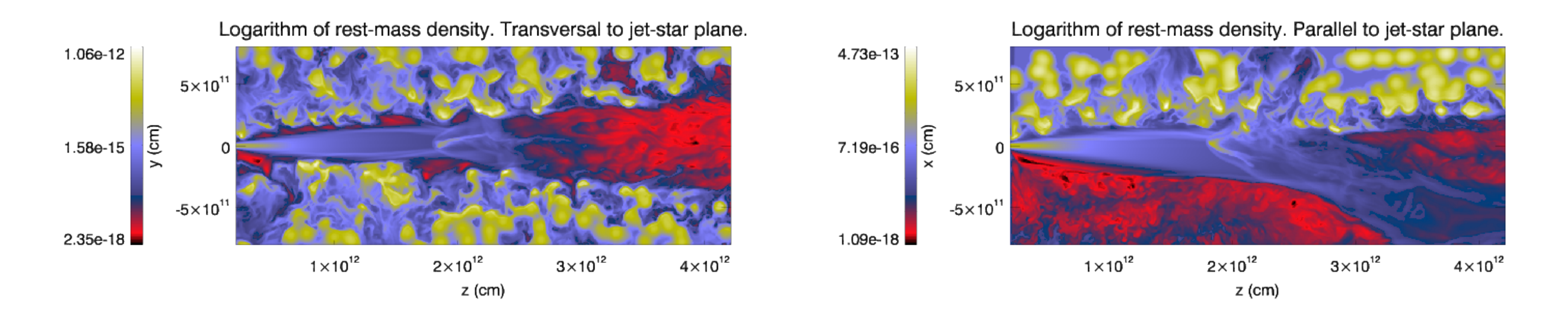}
  \includegraphics[clip,angle=0,width=0.9\textwidth]{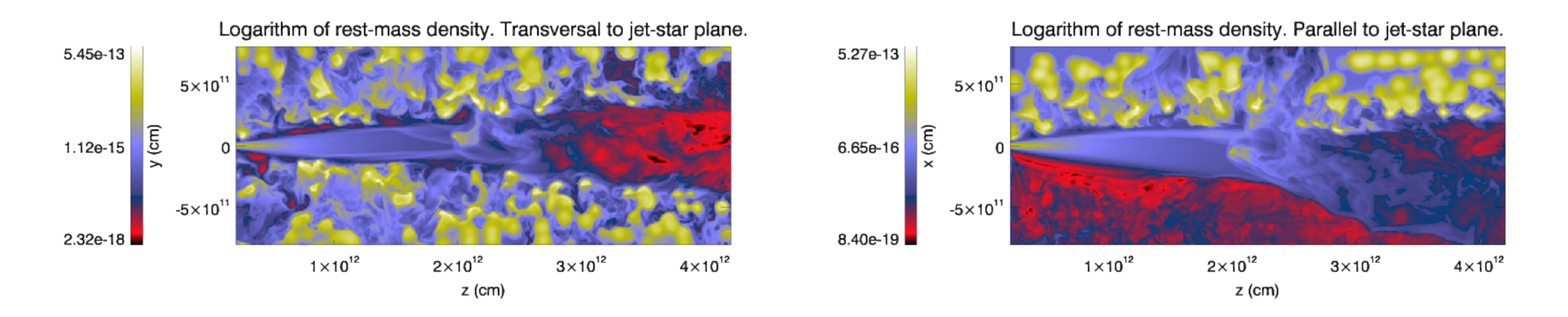}
  \includegraphics[clip,angle=0,width=0.9\textwidth]{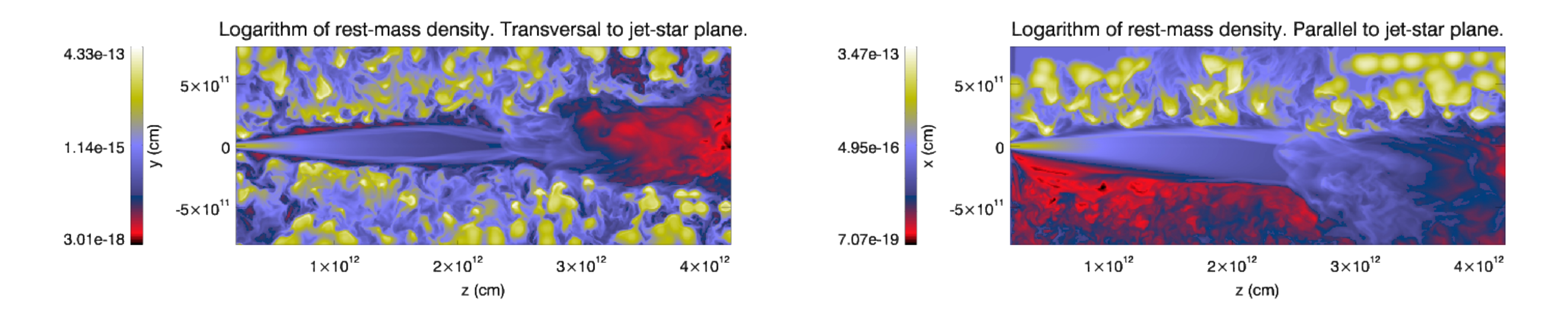}
  \caption{Axial cuts of rest-mass density ($\mathrm{g/cm^3}$) at different instants for jet B.  Top panels stand for $t=470$~s, 
second row for $t=1640$~s, third one for $t=1770$~s and bottom one for the last snapshot of the 
simulation, at $t=1970$~s.}
  \label{fig:maps8}
  \end{figure*}

\begin{figure*}[!t]
    \centering
  \includegraphics[clip,angle=0,width=0.9\textwidth]{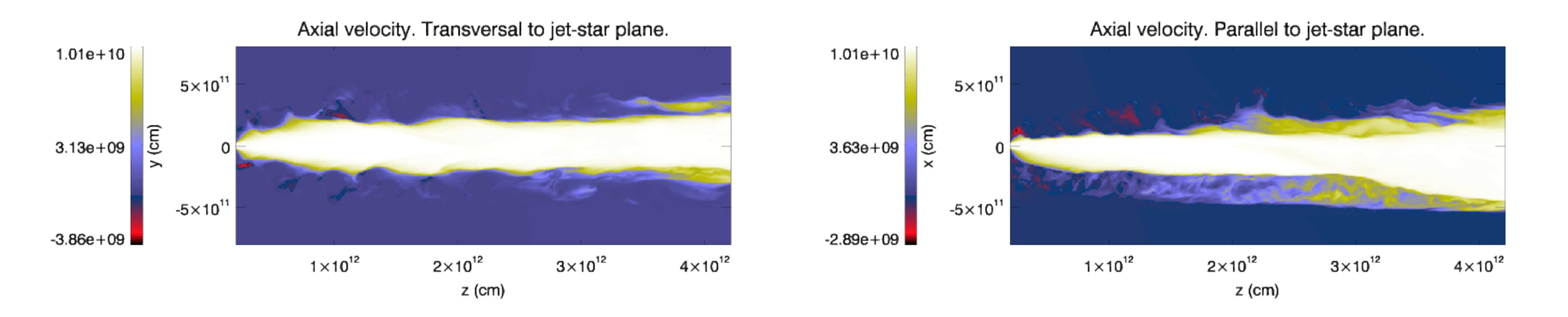}
  \includegraphics[clip,angle=0,width=0.9\textwidth]{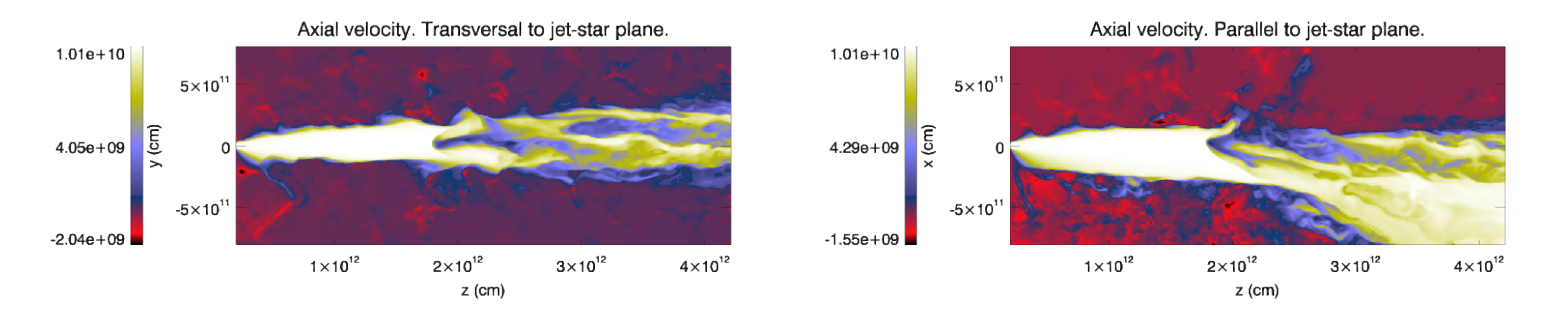}
  \includegraphics[clip,angle=0,width=0.9\textwidth]{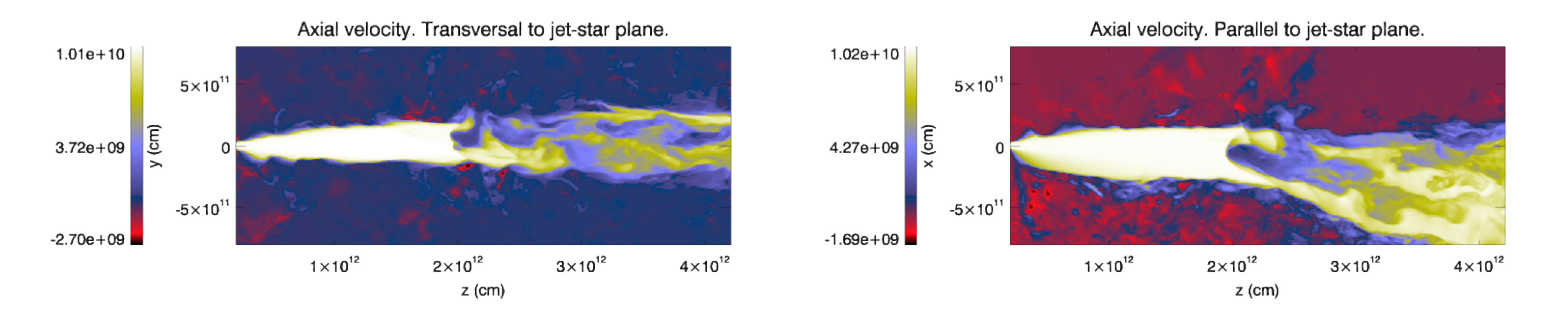}
  \includegraphics[clip,angle=0,width=0.9\textwidth]{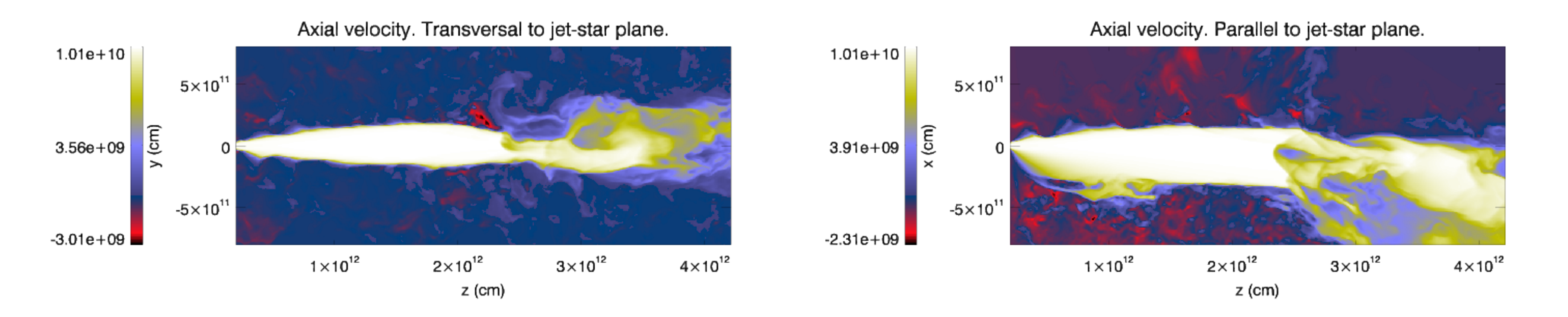}
  \caption{Axial cuts of axial velocity ($\mathrm{cm/s}$) at different instants for jet B. Top panels stand for $t=470$~s, 
second row for $t=1640$~s, third one for $t=1770$~s and bottom one for the last snapshot of the 
simulation, at $t=1970$~s.}
  \label{fig:maps9}
  \end{figure*}

\begin{figure*}[!t]
    \centering
  \includegraphics[clip,angle=0,width=\textwidth]{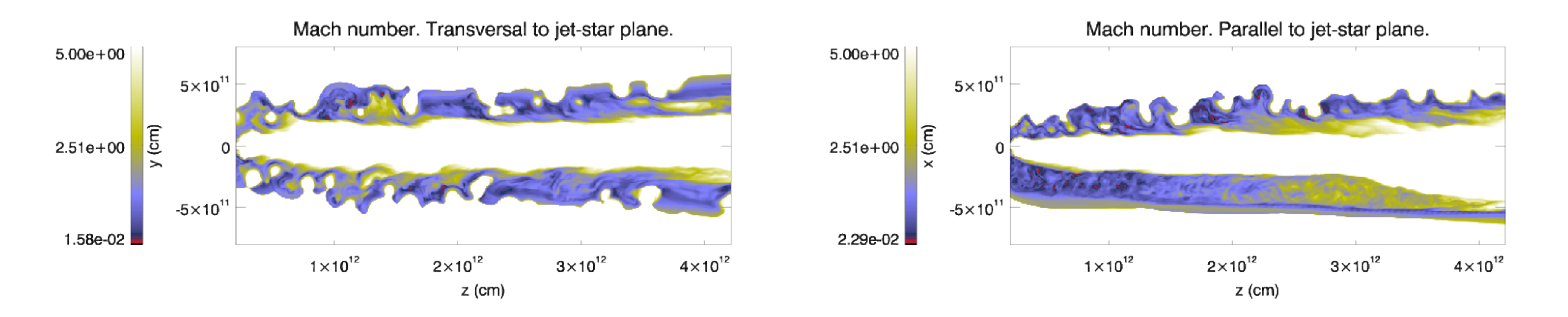}
  \includegraphics[clip,angle=0,width=\textwidth]{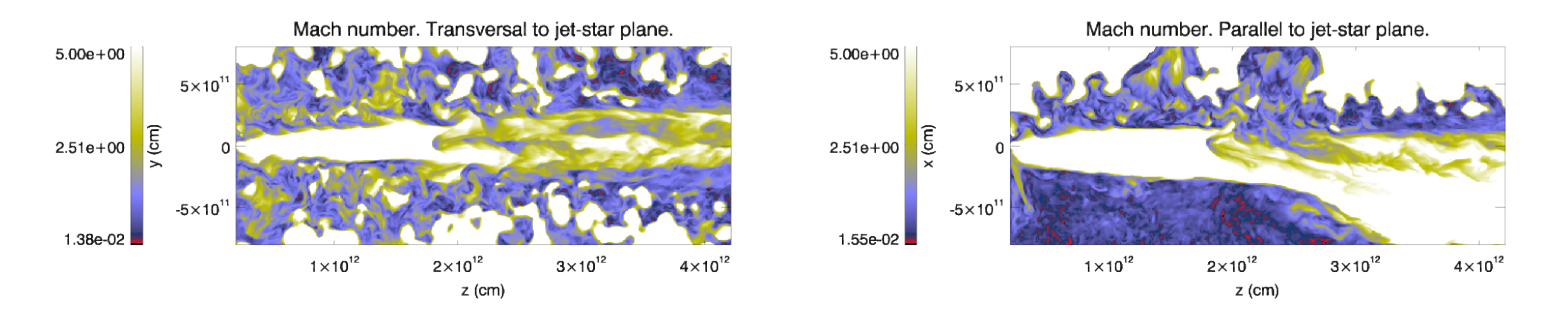}
  \includegraphics[clip,angle=0,width=\textwidth]{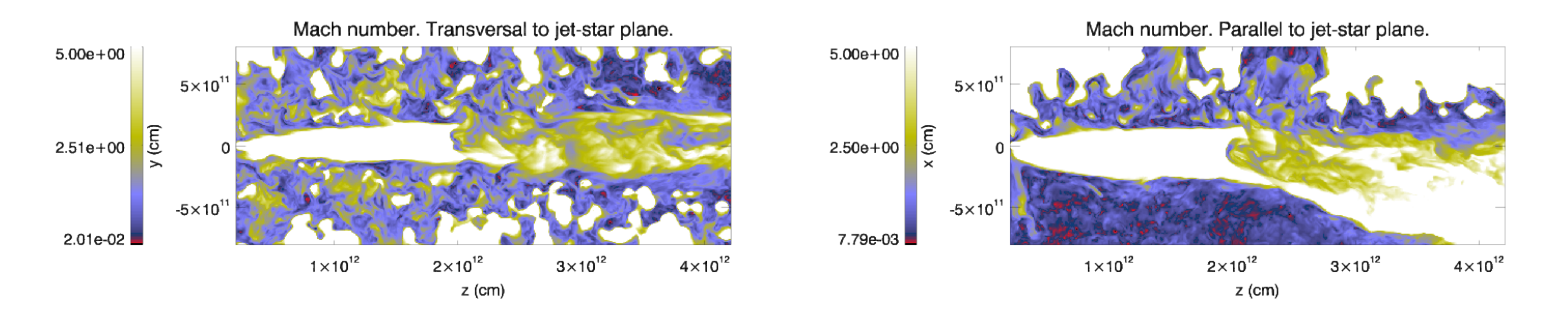}
  \includegraphics[clip,angle=0,width=\textwidth]{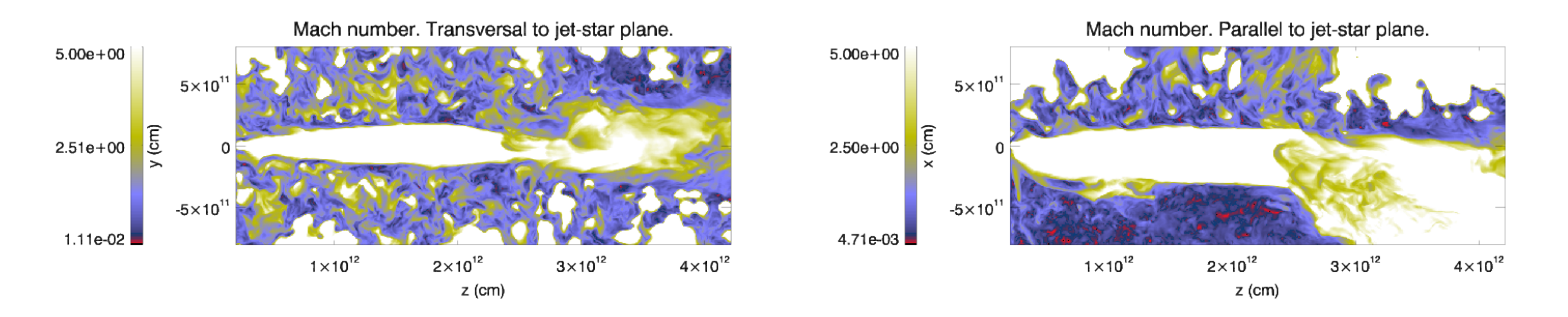}
  \caption{Axial cuts of Mach number at different instants for jet B. Top panels stand for $t=470$~s, second 
row for $t=1640$~s, third one for $t=1770$~s and bottom one for the last snapshot of the 
simulation, at $t=1970$~s. As in Fig.~\ref{fig:maps6}, the color scale is saturated at $M=5$.}
  \label{fig:maps10}
  \end{figure*}

This simulation reproduces the jet-wind interaction during $t \simeq 2000$~seconds for jet~B, with 
$L_{\rm j}=10^{37}$~erg/s. The same qualitative explanation for the
evolution of jet~A applies also to jet B. However, the overpressure of the jet is larger, and the reconfinement 
shock, produced by the global clumpy wind impact, appears further downstream
than in jet~A. There are interesting differences with respect to jet~A due to the entrainement of a clump in 
jet~B at $z\simeq 2.2\times 10^{12}\,\mathrm{cm}$, which is
favored by the strong decrease of the inertia of the jet downstream of the reconfinement shock. This clump 
generates a strong bow shock that eventually covers the whole jet section. Behind the bow shock the
jet is decelerated and a turbulent layer is formed. Once the clump is destroyed, the jet may recover its 
previous configuration, but the constant injection of further inhomogeneities in the wind
makes it unlikely. This event is very illustrative of the clumpy wind disruptive effects. 
Figures~\ref{fig:maps8}, \ref{fig:maps9} and \ref{fig:maps10} show the evolution of jet~B
with time through panels of rest-mass  density, axial velocity and Mach number, respectively. The large 
overpressure of the jet allows some filaments of jet material to propagate between
clumps in the up-wind direction. Again, the jet becomes subsonic downstream of the region where the reconfinement 
shock and the bow shocks generated by the clumps occur
($z\simeq2.4\times10^{12}\,\mathrm{cm}$).

\begin{figure*}[!t]
    \centering
  \includegraphics[clip,angle=0,width=0.9\textwidth]{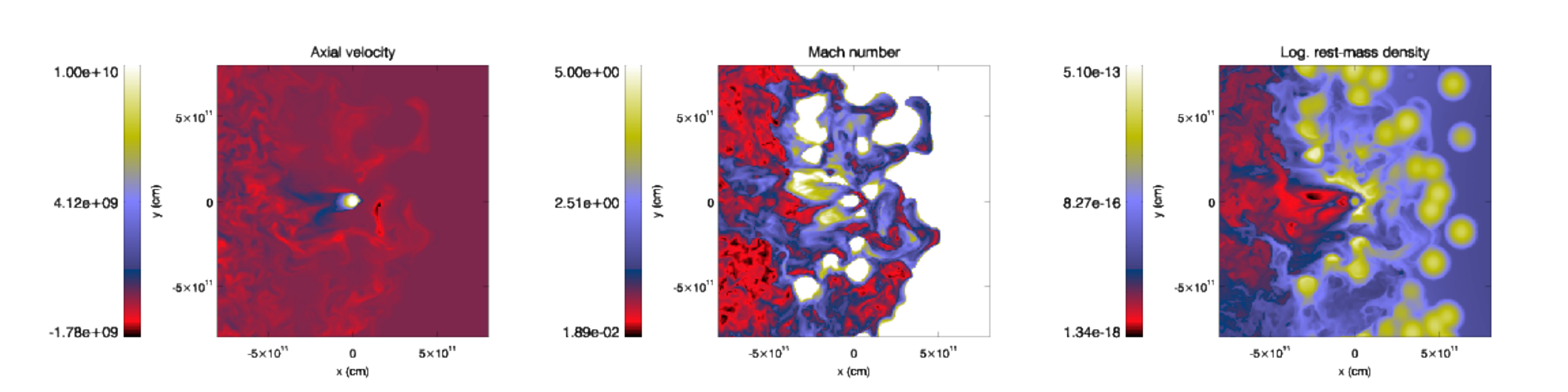}
  \includegraphics[clip,angle=0,width=0.9\textwidth]{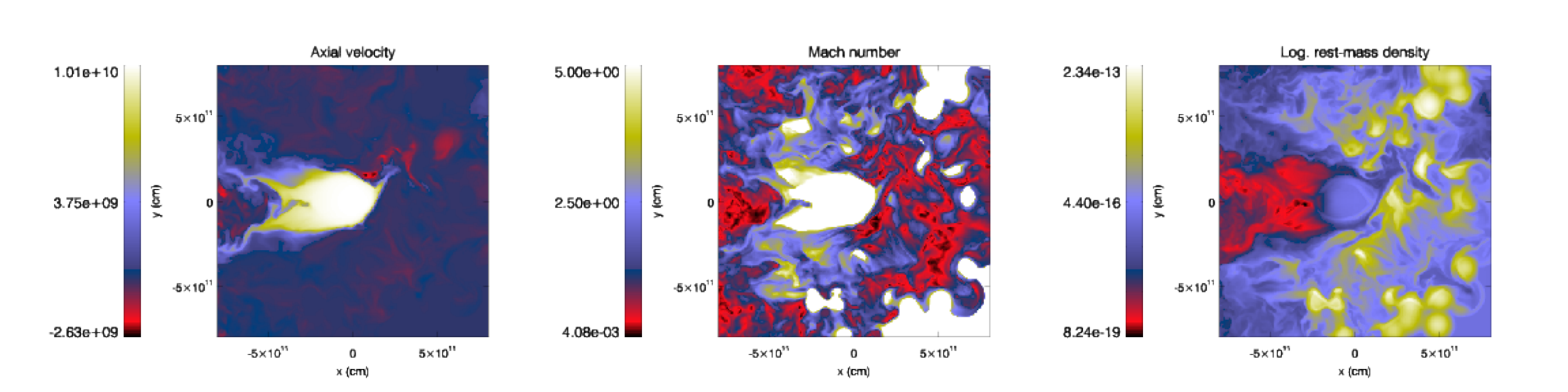}
  \includegraphics[clip,angle=0,width=0.9\textwidth]{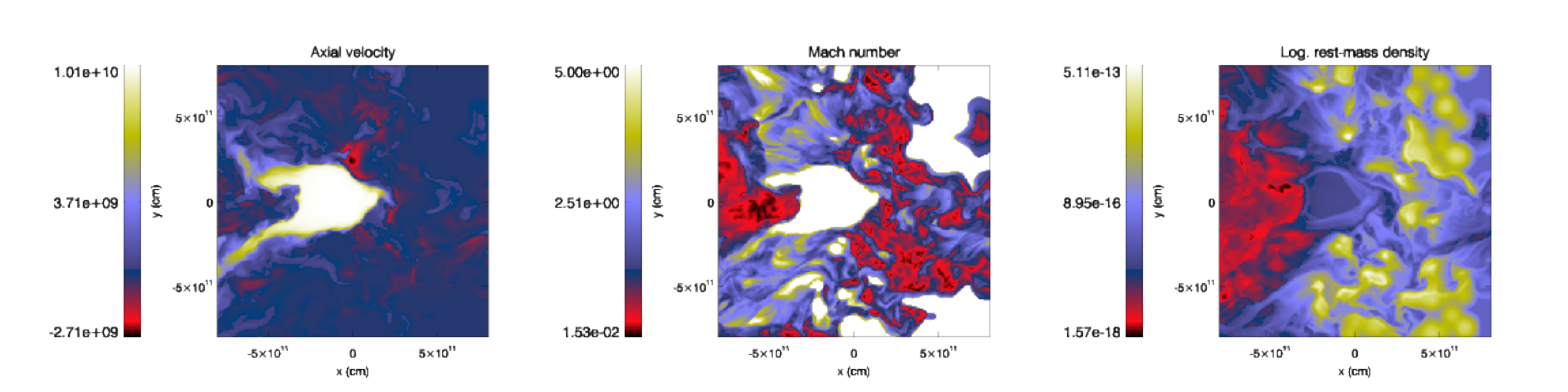}
  \includegraphics[clip,angle=0,width=0.9\textwidth]{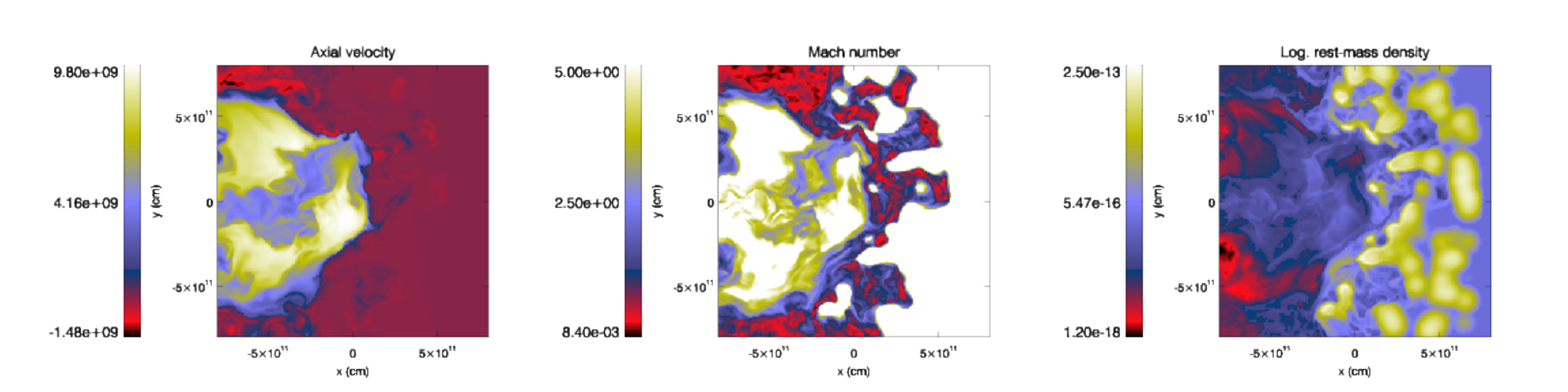}
  \caption{Transversal cuts of rest-mass density ($\mathrm{g/cm^3}$), axial velocity ($\mathrm{cm/s}$) and Mach number for the 
last snapshot ($t=1900$~s) for jet B, at 
$z=0.2,\,1.2,\,2.2,$~and~$3.2\times 10^{12}\,\mathrm{cm}$. The color scale of the Mach number has 
been saturated at $M=5$.}
  \label{fig:maps11}
  \end{figure*}

Figure~\ref{fig:maps11} shows transversal cuts of axial velocity, Mach number and rest-mass density at the last 
snapshot of the simulation at the jet injection position, $z\simeq 2\times
10^{11}\,\mathrm{cm}$, and at $z=1.2,\,2.2,$~and~$3.2\times 10^{12}\,\mathrm{cm}$. At the injection point, the 
image obtained is very similar to that from jet~A (see upper panels in
Fig.~\ref{fig:maps7}). At $z=1.2\times 10^{12}\,\mathrm{cm}$ and $2.2\times 10^{12}\,\mathrm{cm}$, the jet is 
still well collimated with large velocity and Mach number. In these cuts,
filaments of material travelling in the up-wind direction can be observed. The last set of images (bottom row), 
at $z=3.2\times 10^{12}\,\mathrm{cm}$, shows the effect of the entrained
clumps. In the left panel, the traces of the clumps can be seen as low velocity regions, resulting also in 
transonic or even subsonic velocities as indicated by the Mach number maps. The
initial density of the jet at this point was $\rho\simeq 5\times 10^{-17}\,\mathrm{g/cm^{-3}}$ at the start of 
the simulation, whereas the right panel shows densities one order of magnitude
larger within the jet, indicating efficient assimilation of wind material.

\subsection{Jet~A$^\prime$}

\begin{figure*}[!t]
    \centering
  \includegraphics[clip,angle=0,width=0.9\textwidth]{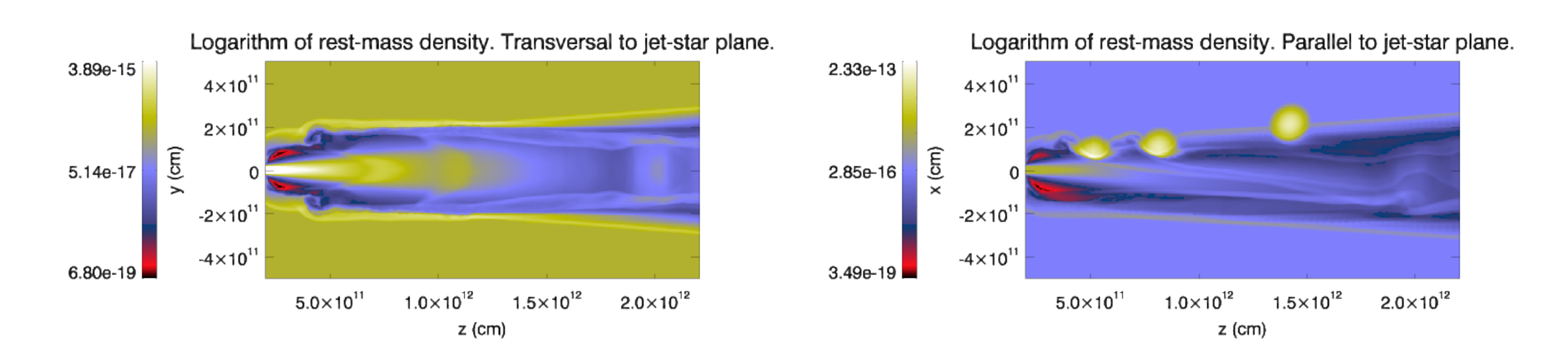}
  \includegraphics[clip,angle=0,width=0.9\textwidth]{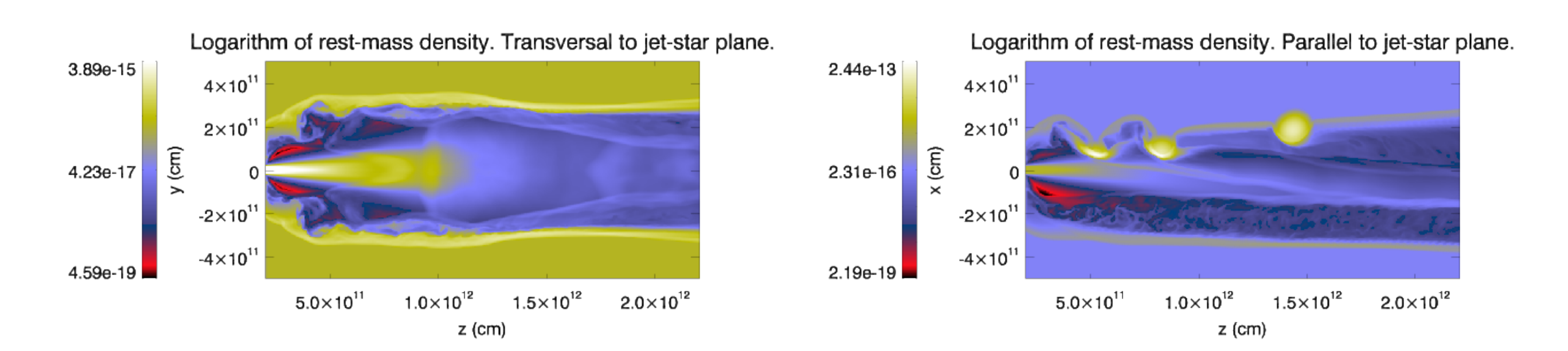}
  \includegraphics[clip,angle=0,width=0.9\textwidth]{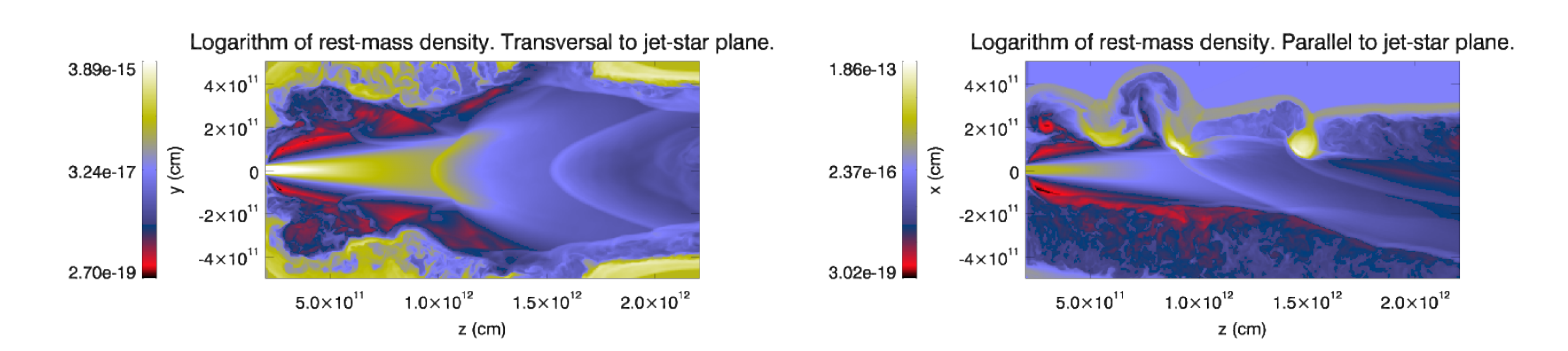}
  \includegraphics[clip,angle=0,width=0.9\textwidth]{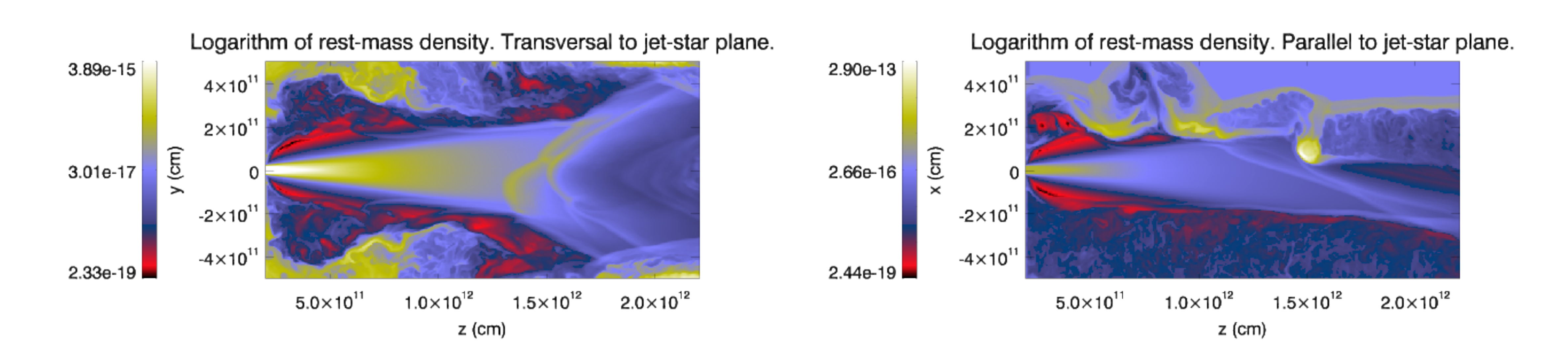}
  \caption{Axial cuts of rest-mass density ($\mathrm{g/cm^3}$) at different instants for jet A'. Top panels stand for 
$t=170$~s, second 
row for $t=320$~s, third one for $t=780$~s, and bottom one for the last snapshot of the simulation, at $t=970$~s.}
  \label{fig:maps12}
  \end{figure*}

\begin{figure*}[!t]
    \centering
  \includegraphics[clip,angle=0,width=\textwidth]{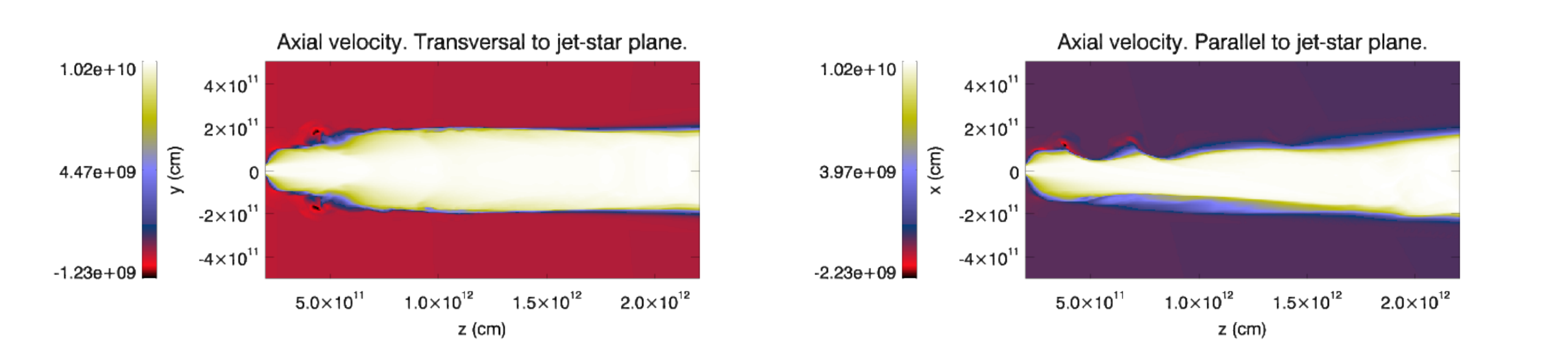}
  \includegraphics[clip,angle=0,width=\textwidth]{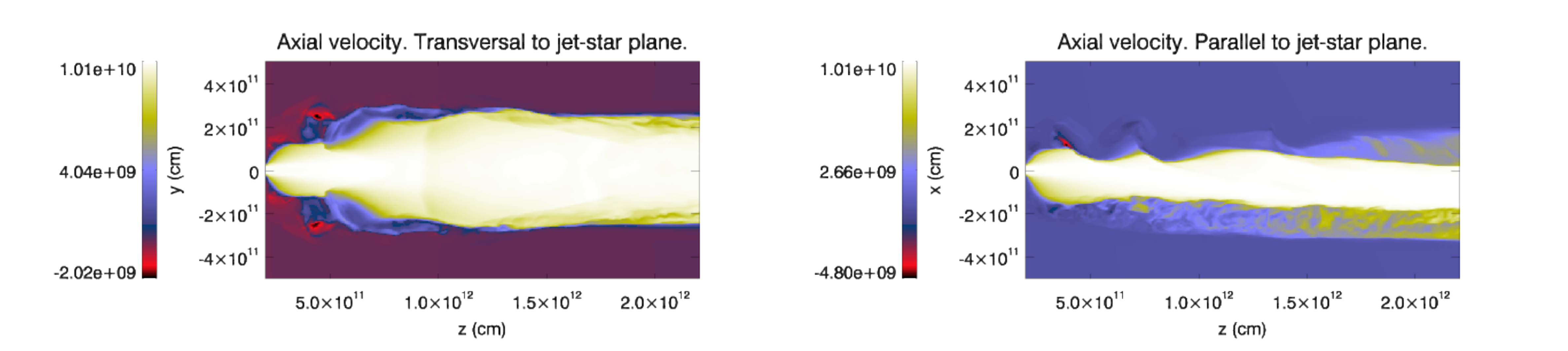}
  \includegraphics[clip,angle=0,width=\textwidth]{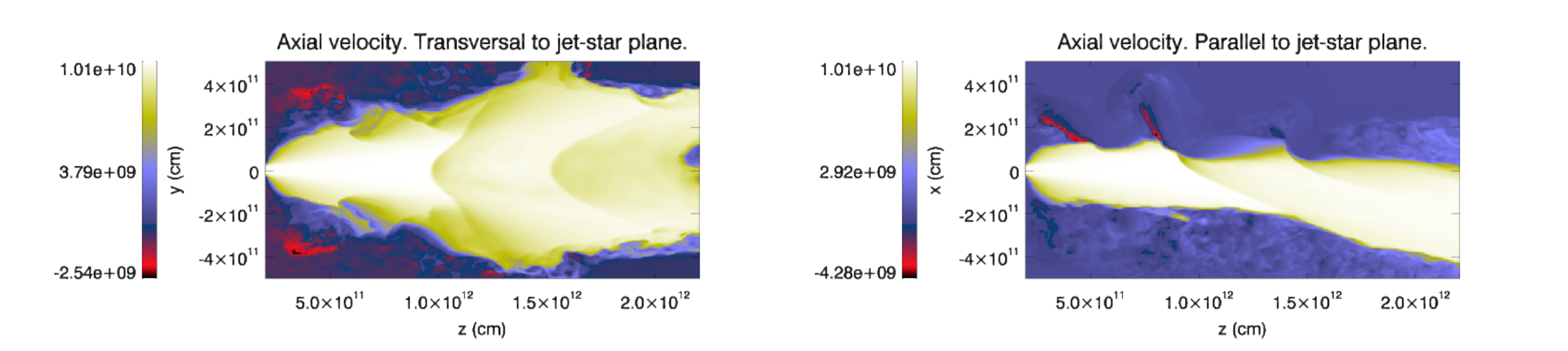}
  \includegraphics[clip,angle=0,width=\textwidth]{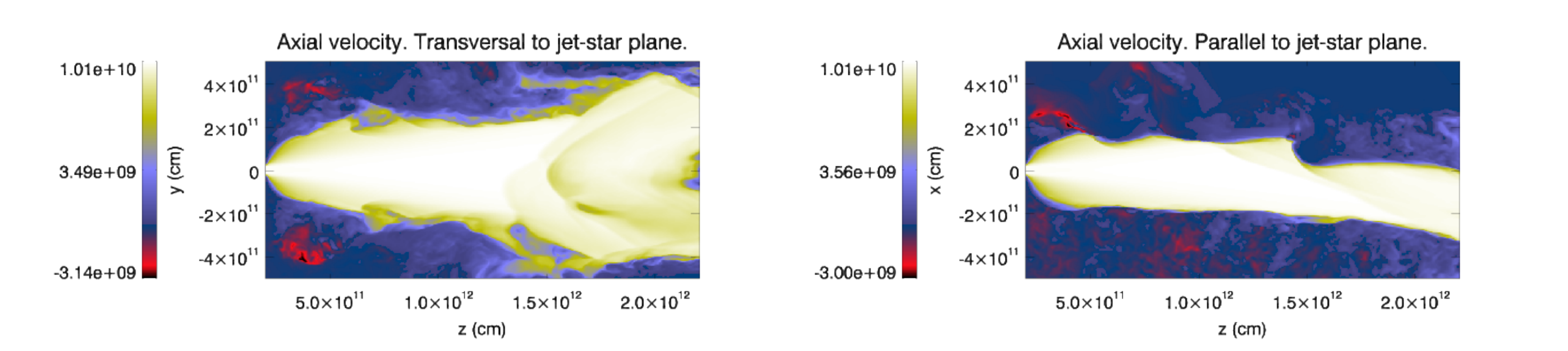}
  \caption{Axial cuts of axial velocity ($\mathrm{cm/s}$) at different instants for jet A'. Top panels stand for $t=170$~s, second 
row for $t=320$~s, third one for $t=780$~s, and bottom one for the last snapshot of the simulation, at $t=970$~s.}
  \label{fig:maps13}
  \end{figure*}

The previous simulations show very complex structures and dynamics. In order to have a clearer idea of the 
processes taking place, we repeated the simulation of jet~A with only three clumps
located initially at $z=0.3,\,0.6$~and~$1.2\times10^{12}\,\mathrm{cm}$, as shown in Fig.~\ref{fig:maps3}. The simulation 
was stopped after $t\simeq 1000$~seconds. Figures~\ref{fig:maps12} and
\ref{fig:maps13} show the time evolution of the interaction of the clumps with the jet. The 
first clump is destroyed in the jet surface, with little dynamical impact on
the jet. Otherwise, the second clump triggers a shock that propagates inside the jet, although the clump itself is 
also destroyed before entering into the jet. Between the first two clumps,
jet material shocked by the second clump rises due to the local pressure gradients 
in the up-wind direction. The third clump shows the same behavior 
as observed in the simulation of jet~B, being able to enter into the jet
before being destroyed, and generating a strong bow shock that causes jet expansion, deceleration and significant 
mass-load.

\begin{figure*}[!t]
    \centering
  \includegraphics[clip,angle=0,width=\textwidth]{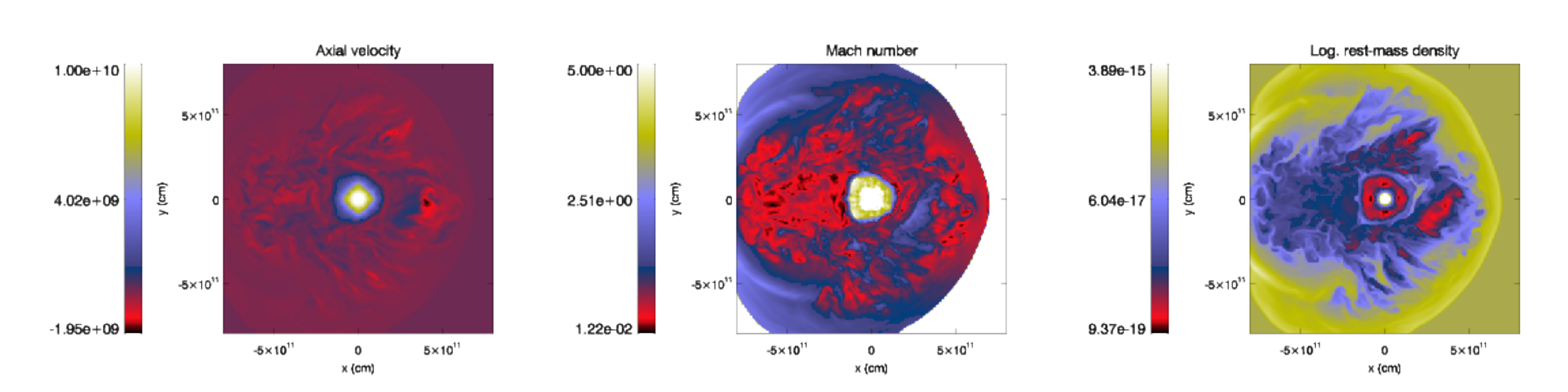}
  \includegraphics[clip,angle=0,width=\textwidth]{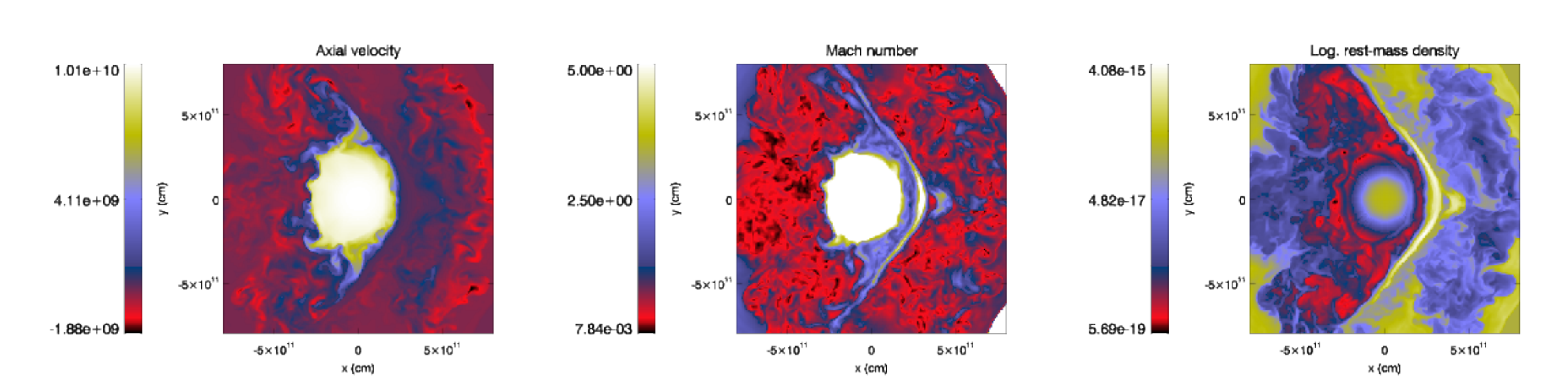}
  \includegraphics[clip,angle=0,width=\textwidth]{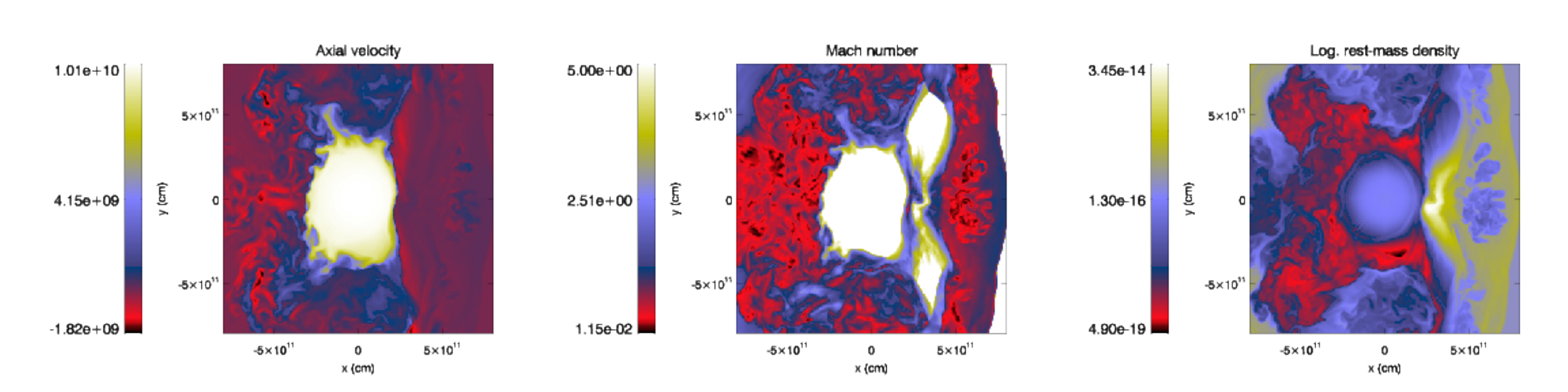}
  \includegraphics[clip,angle=0,width=\textwidth]{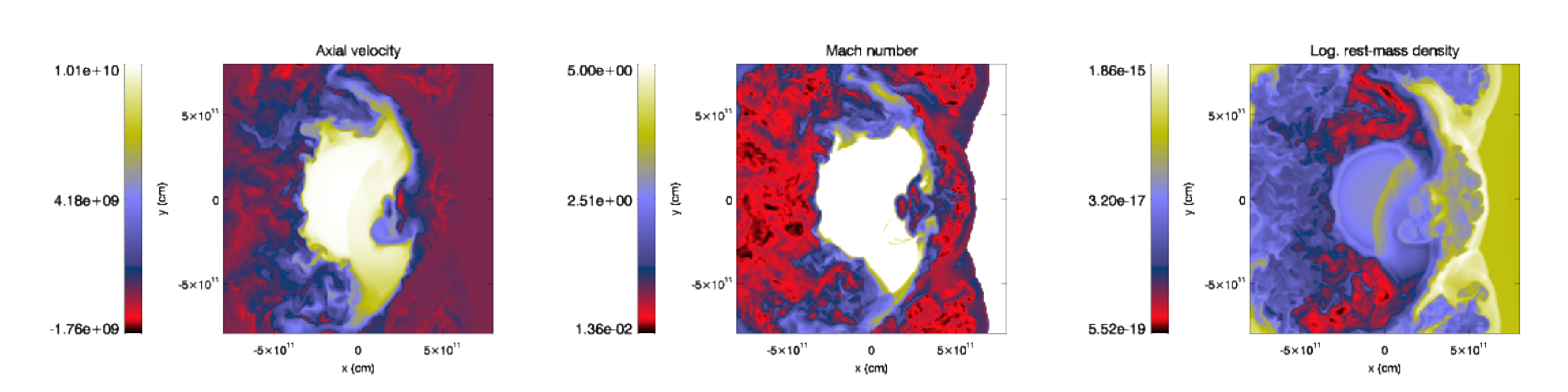}
  \includegraphics[clip,angle=0,width=\textwidth]{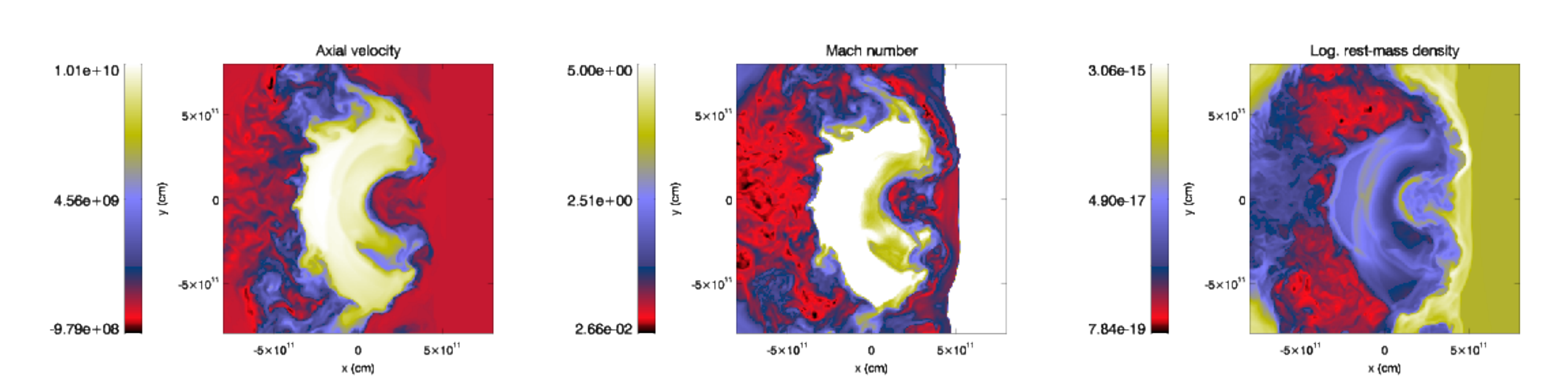}
  \caption{Transversal cuts of rest-mass density ($\mathrm{g/cm^3}$), axial velocity ($\mathrm{cm/s}$) and Mach number for the 
last snapshot ($t=970$~s) of jet A' at 
$z\simeq 0.25,\, 0.75,\, 1,\, 1.5$~and~$1.75\times10^{12}~\mathrm{cm}$. The color scale of the Mach number has 
been saturated at $M=5$.}
  \label{fig:maps15}
  \end{figure*}

Figure~\ref{fig:maps15} shows transversal cuts of axial velocity, Mach number and density (as in 
Figs.~\ref{fig:maps7} and \ref{fig:maps11}). In this case, the cuts are done at $z\simeq
0.25,\, 0.75,\, 1,\, 1.5$~and~$1.75\times10^{12}~\mathrm{cm}$. The structures are much clearer in these plots 
due to the small number of clumps. 
At $z=0.75\times10^{12}~\mathrm{cm}$ we see the disrupted rests of the first clump (originally at
$z=0.3\times10^{12}~\mathrm{cm}$, see bottom panel of Fig.~\ref{fig:maps12}). Initially, this clump crossed the 
shock driven by the jet in the diluted medium, and was disrupted at the jet surface. This interaction already 
implies some loss of initial jet thrust, which is transferred to the clump remains that become part of the jet 
flow. The shock that
this interaction generates is observed as a bow-like structure inside the jet inner boundary in the Mach number 
and density panels at higher values of coordinate $z$. 
The next set of cuts, at
$10^{12}\,\mathrm{cm}$, shows a snapshot of the  disruption of the second clump. This process is here at an earlier 
stage than that of the first clump.  At $z=1.5$~and~$1.75\times10^{12}\,\mathrm{cm}$,
one can see the clear effect of the clump located at $1.4\times10^{12}\,\mathrm{cm}$, which generates a bow-like 
structure entering the jet and a cometary tail, crossed by the transversal cuts, of jet dragged material (see
bottom panels in Figs.~\ref{fig:maps12} and \ref{fig:maps13}). The other bow-like waves, 
visible mainly at $x<0$ in the images, are the bow shocks generated by the interaction with the clumps upstream of 
this position. At $z=1.75\times10^{12}\,\mathrm{cm}$, the jet is dominated by the bow-like structures triggered
by the interactions upstream, and the cometary-like tail from the third clump is seen at $x>0$ region. The 
faster regions of the jet are those that have not been crossed by any of these shocks, i.e., those that lie at 
the smaller values of $x$. The jet keeps a large velocity in those particular regions, but it is in general 
efficiently mass-loaded and decelerated.  

Overall, the picture is similar to that shown in simulations of jets A and B, showing that only a few clumps 
generated by inhomogeneities in the injection of the wind could produce already significant changes in the properties of 
the jet flow.

\section{Discussion}\label{disc}

\subsection{Jet propagation}\label{disc1}

The simulated jets are strongly deviated from their original direction and show disrupted morphologies at the end 
of the calculations. Initially, the first interactions at small values of
$z$ trigger helical patterns that, in the absence of further perturbations, could couple to Kelvin-Helmholtz 
unstable modes. However, the reconfinement shock triggered by the 
ram pressure of the clumpy wind on the jet plus individual interactions with multiple clouds 
downstream of this shock, generate non-linear structures that are most important for the jet evolution.
The position of the reconfinement shock is located within the binary region, as predicted by PB08.
There, the jet is decelerated and heated, thus becoming even weaker relative to the wind thrust, as also shown 
in PB08 and PBK10. In the present case, the clumps from the 
up-wind region can penetrate in the jet already at $z\sim 1-2\times
10^{12}$~cm, generating additional bow-like shocks and decelerating even more the jet flow. The 
deviation of the jet favors the presence of a region, on the down-wind side of the jet, to which, by the
end of simulations, the clump bow shocks and jet mixing have still not reached. In this region, the flow keeps a 
relatively large axial velocity in this direction. However, both shocks and clump-mixed jet
material should eventually fill the whole jet at larger $z$, so the description just presented (see Figs~\ref{fig:maps8}, 
\ref{fig:maps9} and \ref{fig:maps10}) can be extended to the whole jet cross
section. As a  result, farther downstream the jet becomes mass-loaded, slow and transonic. Although a similar 
result was already obtained in the homogeneous wind case (PB08, PBK10), wind
clumping significantly potentiates jet destruction. 


When comparing jet~B in this work with jet~2 in PBK10, it is remarkable that the presence of a forward shock and  the absence of clumps in the latter have a strong influence on the
long-term evolution of the jet. The jet-driven forward shock in the wind and the corresponding cocoon (see PBK10) keep the wind at some distance from the jet. If a homogeneous wind were
already in contact with jet~2 in PBK10, a smooth shear layer would form. Otherwise, the presence of clumps in jet~B enhances  the thrust locally in the wind, increasing mass, momentum and
energy exchange. Jet~2 in PBK10 thus shows a larger  degree of collimation and is just slightly deviated from its original direction of propagation, contrary to what  we observe in the case
of jet~B in this paper. From this work we can conclude that jet luminosities  $L_{\rm j}\gtrsim 10^{37}\,\dot{M}_{-6}\mathrm{erg/s}$ ($\dot{M}_{-6}=\dot{M}/10^{-6}\,M_\odot$/yr) are needed
if the jet is not to be destroyed when crossing the binary system. We notice that these dynamical arguments  favor $L_{\rm j}\gtrsim 10^{37}$ and $10^{38}$~erg/s in the HMMQ Cygnus~X-1 and
Cygnus~X-3, respectively. The cocoon/clumpy wind case deserves few words. This situation takes place when the forward shock is well within the binary system, and the cocoon pressure is
high, preventing clumps from entering the cocoon. When the forward shock has reached the outskirts of the binary, the cocoon pressure drops quickly (PBK10), allowing wind clumps to
penetrate into the cocoon and reach the jet, and eventually dissipate the cocoon away.

\subsection{Clump evolution}\label{disc2}
 
In jets~A and B, those clumps reaching the jet relatively close to its base are destroyed just by jet expansion 
or erosion. These interactions trigger a shock wave that propagates inside the jet, and when interactions are 
frequent such a wave forces the strongly disruptive asymmetric recollimation shock. This phenomenon is 
illustrated in Fig.~\ref{fig:maps12}), in which the first clump is completely destroyed by jet expansion, 
whereas the second one, even when it does not fully penetrate into the jet, triggers a shock strong that 
propagates all through the latter lasting for the whole simulation (several $t_{\rm d}$; see Sect.~\ref{phys}). 
The clump at the highest $z$ in jet~A' ($1.4\times 10^{12}$~cm) is shocked but still not significantly disrupted by 
instabilities after few $t_{\rm d}$. Later, this clump could be destroyed or may eventually escape 
the jet, although jet bending in the down-wind direction makes the latter unlikely. When clumps are disrupted 
inside the jet, all the clump mass is entrained by the flow. The level of mass loading can be easily estimated from 
the amount of clumps entering into the jet per time unit: 
$\dot{N}_{\rm cj}\sim (\eta/4\pi)(3\dot{M}/4\pi R_{\rm c}^3\rho_{\rm c})\approx 0.02$~clump/s 
or $\approx 5\times 10^{17}$~g/s. This is $\sim 3$ times more mass flux
than in the jet. Therefore, jet deceleration due to mass loading can be very efficient, as is most clearly seen 
in the simulation results for jet~B. The implications also apply to
faster and lighter, but equally powerful jets. Even if a strongly relativistic flow is injected at the jet base, 
for similar jet (relativistic) ram pressures clump penetration into, and
mixing with, the jet will also occur. This will mass-load and brake the jet very quickly. 

The qualitative dynamical scenario presented in ABR09 is validated by our numerical 
work in a semi-quantitative way. Remarkably, as hinted in ABR09, the clump interaction with many clumps 
will reduce the jet kinetic luminosity and also its ram pressure. In this way, clumps can actually keep their 
integrity longer, and penetrate farther inside the jet, even for moderate $f$-values.

\subsection{Radiation considerations}\label{disc3}

In ABR09, the radiation produced by a clump inside a HMMQ jet was calculated. Given the compactness of the 
considered region, i.e. the bow shock formed around the clump, radio
emission was negligible. The luminosities of synchrotron X-ray, and (stellar photon) IC gamma-rays, dominated the non-thermal output, 
with their values anticorrelating depending on the magnetic field. Based on
that work, and extending the study to the multi-clump/jet interaction case simulated here, we qualitatively discuss in what follows the 
expected emission.

The typical number of clumps inside the jet, at the binary scales, can be estimated as $N_{\rm cj}\sim$ few 
times $t_{\rm d}\,N_{\rm cj}$, i.e. $\gtrsim 10$ for the wind and jet properties
adopted in this work (recall that the real clump lifetime is longer than $t_{\rm d}$). Since 
$t_{\rm d}\sim 100\,{\rm s}<R_{\rm j}/v_{\rm w}\sim 1000$~s, clump destruction will likely limit
$N_{\rm j}$ instead of jet-crossing escape. Although the jet can be quite disrupted and mass-loaded at the borders of the system, 
one can derive an estimate for the non-thermal emission, which depending on $B^2/8\pi
u_*$ ($u_*=L_*/4\pi d^2 c$) will be released either through synchrotron (keV-MeV) or IC (GeV-TeV). We note that, 
for $B$-values well below equipartition with the non-thermal electron energy density, synchrotron self-Compton may also contribute to the IC component, although
we neglect this component here for analysis simplicity. We also do not consider non-radiative losses or hadronic 
processes, but refer to ABR09.

Assuming a 10\% efficiency for the kinetic-to-non-thermal energy transfer in clump bow shocks, and accounting 
for the clump-jet covering fraction $\xi_{\rm cj}\sim (R_{\rm c}/\eta d)^2$,
the high-energy emission can reach luminosities of 
$\sim 0.1\,N_{\rm cj}\,\xi_{\rm cj}\,L_{\rm j}\sim 0.01\,L_{\rm j}$. This value may lead to a source that may 
be detectable in GeV
($B^2/4\pi u_*<1$) and even in TeV ($B^2/4\pi u_*\ll 1$), showing persistent plus fluctuating variability 
components \citep{owo09}. Strong TeV photon absorption may be avoided since
most of the emission would come from the borders of the binary \citep[e.g.,][]{bos09}. For 
$B^2/4\pi u_*\gtrsim 1$, synchrotron soft gamma-rays would dominate the non-thermal output. 
Thermal radiation, peaking at several keV, from the shocked clumps may be also present (ABR09). but it is minor
for the adopted clump properties.

It is noteworthy that for a steep clump-size distribution (see Sect.~\ref{phys}), i.e. with most of the 
clump-jet interaction occurring in a quasi-homogeneous wind regime, still several big
clumps could enter the jet. This would likely lead to a lower level of persistent emission, but time to time 
chance would bring few big and dense clumps together inside the jet, rendering
sudden and bright events with durations of a few 
$1000\,(R_{\rm c}/10^{11}\,{\rm cm})\,(c_{\rm c}/10^8\,{\rm cm})^{-1}$~s (ABR09). 
These statistically
expected violent events could be related to a temporary complete destruction of the jet for large enough clumps.

Although the jets of microquasars present a complex phenomenology of their own \citep{fen04,fen09}, on top of that  it is expected additional phenomena linked to the jet-wind interaction in
HMMQ, as this and previous works (PB08, PBK10) show. Since the jet power is strongly linked to the mass-loss rate,  HMMQ jets detectable in X-rays or gamma-rays will be probably found in
systems with moderate-to-strong winds. Therefore, unless particle acceleration is negligible in clump bow shocks, HMMQ  phenomenology at high-energies must be strongly affected by them,
showing wind-related strong variability at high energies unless the wind is quite homogeneous or clumps with  $R_{\rm c}>10^{10}$~cm are completely missing. Concerning (powerful)
transient ejections, usually associated to X-ray  state transitions, we note that such an ejections will require some time to form. If this takes hours, the wind will have  time to surround
the transient jet. Then, the clump impact will be as described here unless the jet is too powerful. If powerful blobs would appear as discrete even at the scales of the binary, they may
have too much inertia to be significantly affected by the wind.

The dynamical impact of the stellar wind on the jet, enhanced by clumpiness, should not be neglected when 
interpreting radio emission from HMMQ. Even if jets escape from
disruption, the enhanced jet entropy and bending, even by small angles ($\sim 10^\circ$), could have 
observational consequences. The reason is that bending pushes jet material farther from
the orbital plane axis, increasing the strength of the Coriolis force exerted by the stellar wind and the orbital 
motion. The farther away from this axis the jet reaches, the stronger this
force gets, enhancing jet heating, turbulence and bending. Although quantitative predictions call for a detailed 
study, this region may be observationally probed using VLBI techniques. Far
enough from the binary, once the jet has become too wide to be affected by the orbital motion, a collimated 
supersonic flow may form again. If the energy and momentum fluxes kept enough
anisotropy after crossing the system and suffer from orbital motion, 
given a negligible external pressure a jet-like structure could form again and propagate unstopped 
up to pc scales, terminating in the ISM \citep[e.g.][]{bor09,bos11,yoo11}. Despite the different
outflow geometry, similar phenomena are also expected in high-mass binaries hosting 
non-accreting pulsars \citep[see][]{bos11b}.

\begin{acknowledgements}
The authors want to thank the anonymous referee for very useful 
and constructive comments and suggestions.
MP 
acknowledges
support by the Spanish ``Ministerio de Ciencia e Innovaci\'on''
(MICINN) grants AYA2010-21322-C03-01, AYA2010-21097-C03-01 and
CONSOLIDER2007-00050.
The research leading to these results has received funding from the European
Union Seventh Framework Program (FP7/2007-2013) under grant agreement
PIEF-GA-2009-252463. V.B.-R. acknowledges support by the Spanish 
Ministerio de Ciencia e Innovaci\'on
(MICINN) under grants AYA2010-21782-C03-01 and FPA2010-22056-C06-02.  
We acknowledge the Spanish Supercomputing
Network for the computational time allocated for the simulations. The 
simulations were performed in Mare Nostrum,
at the Barcelona Supercomputing Centre.

\end{acknowledgements}
\bibliographystyle{aa}

\begin{thebibliography}{}
\small{
\bibitem[Abdo et al.(2009)]{abd09} Abdo, A.~A., et al.\ 2009, Science, 326, 1512
\bibitem[Albert et al.(2007)]{alb07} Albert, J., Aliu, E., Anderhub, H., et al. 2007, ApJ, 665, L51
\bibitem[Araudo et al.(2009)]{ara09} Araudo, A., Bosch-Ramon, V., Romero, G.~E., 2009, A\&A, 503, 673 (ABR09)
\bibitem[Araudo et al.(2011)]{ara11} Araudo, A.~T., Bosch-Ramon, V., Romero, G.~E. Proceedings of the 25th Texas Symposium on Relativistic Astrophysics - TEXAS 2010, Heidelberg, Germany [astro-ph/1104.1730]
\bibitem[Barkov et al.(2010)]{bar10}Barkov, M., Aharonian, F.~A., Bosch-Ramon, V. 2010, \apj, 724, 1517 
\bibitem[Blandford \& Znajek(1977)]{bla77} Blandford, R. D. \& Znajek, R. L. 1977, MNRAS, 179, 433    
\bibitem[Blandford~\&~Koenigl(1979)]{bla79} Blandford, R. D. \& Koenigl, A. 1979, ApL, 20, 15 
\bibitem[Blandford \& Payne(1982)]{bla82} Blandford, R. D. \& Payne, D. G. 1982, MNRAS, 199, 883
\bibitem[Barkov \& Khangulyan(2011)]{bar11} Barkov, M. V. \& Khangulyan, D. 2011, MNRAS, submitted [astro-ph/1109.5810]
\bibitem[Bordas et al.(2009)]{bor09} Bordas, P., Bosch-Ramon, V., Paredes, J. M., Perucho, M. 2009, A\&A, 497, 325
\bibitem[Bosch-Ramon \& Khangulyan(2009)]{bos09} Bosch-Ramon, V. and Khangulyan, D., Int. Journ. Mod. Phys. D 2009, 18, 347
\bibitem[Bosch-Ramon \& Barkov(2011)]{bos11b} Bosch-Ramon, V. \& Barkov, M. V. 2011, A\&A, in press [astro-ph/1105.6236]
\bibitem[Bosch-Ramon et al.(2011)]{bos11} Bosch-Ramon, V., Perucho, M., Bordas, P. 2011, A\&A, 528, 89
\bibitem[Fender et al.(2004)]{fen04} Fender, R.~P., Belloni, T.~M., Gallo, E. 2004, MNRAS, 355, 1105
\bibitem[Fender et al.(2009)]{fen09} Fender, R. P., Homan, J., Belloni, T. M. 2009, MNRAS, 396, 1370 
\bibitem[Klein et al.(1994)]{kle94} Klein, R.~I., McKee, C.~F., Colella, P. 1994, ApJ, 420, 213 
\bibitem[Komissarov et al.(2007)]{kom07} Komissarov, S. S.; Barkov, M. V., Vlahakis, N., K\"onigl, A. 
\bibitem[Jones et al.(1996)]{jon96} Jones, T.~W., Ryu, D., Tregillis, I. L. 1996, ApJ, 473, 365 
\bibitem[Mirabel et al.(1999)]{mir99} Mirabel, I.~F. \& Rodr\'iguez, L. F. 1999, ARA\&A 37, 409
\bibitem[Moffat(2008)]{mof08} Moffat, A. F. J. 2008, Proceedings of an international workshop: Clumping in hot-star winds, Potsdam, Germany, Hamann, Wolf-Rainer (ed.) ; Feldmeier, Achim (ed.) ; Oskinova, Lidia M. (ed.). ISBN 978-3-940793-33-1., p.17 
\bibitem[Myasnikov et al.(1998)]{mya98} Myasnikov, A. V., Zhekov, S. A., Belov, N. A. 1998, MNRAS, 298, 1021 
\bibitem[Owocki \& Cohen(2006)]{owo06} Owocki, S.P., Cohen, D.H. 2006, ApJ, 648, 565
\bibitem[Owocki et al.(2009)]{owo09} Owocki, S. P., Romero, G. E., Townsend, R. H. D., Araudo, A. T. 2009, ApJ, 696, 690
\bibitem[Perucho \& Bosch-Ramon(2008)]{per08} Perucho, M., Bosch-Ramon, V. 2008, A\&A, 482, 917 (PB08)  
\bibitem[Perucho et al.(2010a)]{per10} Perucho, M., Bosch-Ramon, V., Khangulyan, D. 2010a, A\&A, 512, L4 (PBK10)
\bibitem[Perucho et al.(2010b)]{pe10} Perucho, M., Mart\'{\i}, J.~M., Cela, J.~M., Hanasz, M., de La Cruz, R., 
Rubio, F. 2010b, A\&A, 519A, 41
\bibitem[Pittard et al.(2010)]{pit10} Pittard, J. M., Hartquist, T. W., Falle, S. A. E. G. 2010, MNRAS, 405, 821 
\bibitem[Rib\'o(2005)]{rib05} Rib\'o, M. in Future Directions in High Resolution Astronomy: The 10th Anniversary of the VLBA, (ASPC, 2005) 340, 421 [astro-ph/0402134]
\bibitem[Romero et al.(2010)]{rom10} Romero, G. E., Del Valle, M. V., Orellana, M. 2010, A\&A, 518, 12
\bibitem[Shin et al.(2008)]{shi08} Shin, M.-S., Stone, J.~M., Snyder, G.~F., 2008, ApJ, 680, 336
\bibitem[Sabatini et al.(2010)]{sab10} Sabatini, S. et al. 2010, ApJ, 712, L10
\bibitem[Tavani et al.(2009)]{tav09} Tavani, M., et al. 2009, Nature, 462, 620
\bibitem[Yoon et al.(2011)]{yoo11} Yoon, D., Morsony, B., Heinz, S., et al. 2011, ApJ, in press [astro-ph/1108.4058]
}
\end{thebibliography}

\end{document}